\documentclass[useAMS,usenatbib]{mn2e}

\usepackage{amssymb}
\usepackage{graphicx}
\usepackage{natbib}
\usepackage{lscape}
\usepackage{rotating}
\usepackage{txfonts}

\newcommand{\kms}{km\ s$^{-1}$}

\newcommand{\ha}{H$\alpha$}

\newcommand{\loglxlb}{$\log[L_{\rm X}/L_{\rm BOL}]$}
\newcommand{\cz}{\ensuremath{C_Z}}
\newcommand{\pv}{\ensuremath{P_V}}
\newcommand{\nv}{\ensuremath{N_V}}

\newcommand{\bz}{\ensuremath{\langle B_z\rangle}}

\newcommand{\nz}{\ensuremath{\langle N_z\rangle}}

\newcommand{\fo}{\ensuremath{f^\parallel}}
\newcommand{\fe}{\ensuremath{f^\perp}}

\title[Magnetometry of a sample of massive stars in Carina]{Magnetometry of a sample of massive stars in Carina \thanks{Based on data collected at ESO under Program ID 386.D-0624A.}}
\author[Y. Naz\'e et al. ]{\parbox{\textwidth}{Ya\"el Naz\'e$^{1}$\thanks{E-mail:
naze@astro.ulg.ac.be; Research Associate FRS-FNRS}, Stefano Bagnulo$^{2}$, V\'eronique Petit$^{3}$, Thomas Rivinius$^{4}$, Gregg Wade$^{5}$, Gregor Rauw$^{1}$, Marc Gagn\'e$^{3}$} \vspace{0.4cm} \\
$^{1}$ GAPHE, D\'epartement AGO, Universit\'e de Li\`ege, All\'ee du 6 Ao\^ut 17, Bat. B5C, B4000-Li\`ege, Belgium\\
$^{2}$ Armagh Observatory, College Hill, Armagh BT61 9DG, Northern Ireland, UK\\
$^{3}$ Department of Geology \& Astronomy, West Chester University, West Chester, PA 19383, USA\\
$^{4}$ ESO - European Organisation for Astronomical Research in the Southern Hemisphere, Casilla 19001, Santiago 19, Chile\\
$^{5}$ Department of Physics, Royal Military College of Canada, PO Box 17000, Station Forces, Kingston, ON K7K 4B4, Canada}
\begin{document}

%\date{Accepted 1988 December 15. Received 1988 December 14; in original form 1988 October 11}

\pagerange{\pageref{firstpage}--\pageref{lastpage}} \pubyear{2002}

\maketitle

\label{firstpage}

\begin{abstract}
X-ray surveys of the Carina nebula have revealed a few hard and luminous sources associated with early-type stars. Such unusual characteristics for the high-energy emission may be related to magnetically-confined winds. To search for the presence of magnetic fields in these objects, we performed a limited spectropolarimetric survey using the FORS instrument. The multi-object mode was used, so that a total of 21 OB stars could be investigated during a one-night-long run. A magnetic field was detected in two objects of the sample, with a 6$\sigma$ significance; Tr16-22 and 13. Such a detection was expected for Tr16-22, as its X-ray emission is too bright, variable and hard, compared to other late-type O or O+OB systems. It is more surprising for Tr16-13, a poorly known star which so far has never shown any peculiar characteristics. Subsequent monitoring is now needed to ascertain the physical properties of these objects and enable a full modelling of their magnetic atmospheres and winds.
\end{abstract}


\section{Introduction}
Often invoked in the past in case of unexplained peculiarities, the role of magnetic fields has long been a subject of debate in the massive star community. Indeed, on the theoretical side, these stars lack the convective envelopes thought to be responsible for the magnetic dynamo in late-type stars whereas, on the observational side, line dilution due to bright companions, spectral contamination by emission, and the relative rarity and breadth of spectral lines limit the detectability of magnetic Zeeman signatures in hot stars. At the beginning of this century, only one half-dozen hot stars (i.e. with spectral types of at least B2) had been confidently detected to be magnetic; the earliest of this sample was only of spectral type B1.5. In the last decade, however, the situation has improved dramatically thanks to the advent of efficient spectropolarimeters attached to mid- and large-size telescopes (such as e.g. FORS at the ESO VLT or ESPaDOnS at the CFHT). These instruments led to the detection of magnetic fields in about 30 hot stars (for a few recent cases: \citealt{riv10,hub11,pet11,gru12}).

The role of magnetic fields in massive stars has thus been re-assessed. While \citet{mae03} focused on understanding their effect on the structure and evolution of massive stars, \citet{bab97a} had unveiled their impact on the stellar winds and therefore on the X-ray emission of massive stars. In a seminal paper, they proposed a semi-analytic model for magnetically channeled wind shocks wherein a strong dipole magnetic field guides the stellar winds from two opposite hemispheres towards the magnetic equator where the flows collide: the gas is heated up to high temperatures, leading to hard X-ray emission. Such channelling can naturally lead to a shock-heated equatorial region. 

This model was first developed and applied to explain the peculiar X-ray emission of the magnetic and chemically peculiar Bp star IQ\,Aur. Following the work of \citet{sta96}, it was extended to explain the peculiarities of the O7 star $\theta^1$\,Ori\,C \citep{bab97b}. Self-consistent numerical magnetohydrodynamic models by \citet{udd02} further explored the wind-field interaction in hot stars, ultimately establishing $\theta^1$\,Ori\,C as a prototype for which the 15d modulations of the \ha, UV, and X-ray emissions are well explained by a magnetic oblique rotator model. Furthermore, \citet{udd06} have shown that in the presence of significant stellar rotation, centrifugal support can lead to a stable accumulation of material near the magnetic equator. The best example here is the Bp star $\sigma$\,Ori\,E, for which the magnetic energy is so strong compared to the stellar wind energy that the magnetosphere extends to many stellar radii \citep{tow05}. For both cases, the X-ray emission appears very peculiar, as it is altogether bright and hard. Indeed, ``normal'' O-type stars have relatively soft X-ray emission ($kT\sim0.2-0.7$\,keV) of a well-defined strength (the so-called canonical relation, \loglxlb $\sim10^{-7}$, see \citealt{naz11,naz09} and references therein). Both $\theta^1$\,Ori\,C and $\sigma$\,Ori\,E therefore seem to comply reasonably well with model expectations for magnetically confined winds. 

It must however be noted that, while bright and hard X-ray emission may be considered as a good hint for the presence of magnetic fields, the contrary is not necessarily true. Indeed, some magnetic objects display soft X-ray emission: this is notably the case of Of?p stars (see e.g. \citealt{naz10,naz12}) and some B stars (see \citealt{gud09,osk11} and references therein). While the overall X-ray luminosity of these objects follows rather well the expected values, it is clearly not the case of their plasma temperature \citep{osk11}, a fact which currently remains unexplained.

The Chandra X-ray observatory has recently performed a large-scale survey of the Carina nebula \citep[CCCP, see ][]{tow11}. This led to the detection of $>$14\,000 X-ray sources \citep{bro11}, among which a majority of PMS objects and $>$100 massive, hot stars \citep{naz11}. Amongst the latter ones, we selected those that display unusually hard and bright X-ray emission, hence were good candidates for harbouring magnetically-confined winds. Using FORS, we perform a small spectropolarimetric survey of these objects, with the aim of finding new cases of magnetically-confined winds. This paper reports on the results of this campaign. It first presents the observations (Sect. 2), then the results (Sect. 3), to finally summarize and conclude in Sect. 4.

\section{Observations}

\subsection{Observational Strategy}

A total of eight fields in the Carina nebula were observed in spectropolarimetric mode with the FORS2 instrument \citep{app98} of the ESO VLT during a visitor run in March 2011. The choice of the main targets was initially made on the basis of their X-ray emission \citep[CCCP, ][and references therein]{naz11}. We selected bright\footnote{$V<11.$\,mag} early-type stars with high X-ray luminosity (Coll228-68, HD\,93250, Tr16-5, 11, 22 and 64), with a strong hard X-ray component (HD\,93250, HD\,93501, LS\,1865, Coll228-68, Tr16-5, 10, 11, 22 and 64) or with known long-term X-ray variability (HD\,93250, Tr16-22). Those characteristics are compatible with those expected from magnetically-confined winds, explaining our choice for this investigation.

Four fields studied only a single target, because of the lack of nearby massive stars with similar magnitudes: HD\,93250, HD\,93501, HD\,93190, and Tr16-22. The four other fields were taken in multi-object mode, since such adequate neighbours existed: the first one contains Coll228-66, Coll228-68, and HD\,93097, the second one Tr16-10, 11 and 14, the third one LS\,1865, LS\,1853, HD\,305524, and HD\,305534, and the last one Tr16-2, 4, 5, 13, 15, 64, and 115.  The 11 additional objects are field stars which do not show particularly exceptional X-ray characteristics.

In total, 21 OB stars were thus investigated. Our observing run was allocated 10\,hours, that were distributed over three consecutive nights (2011  March 12--14). This allowed us to observe about half of our targets on two different nights. 
%Note that the observations occurred over 3 consecutive nights (2011 March 12--14), to be able to duplicate the exposures with the aim of confirming the results, as much as possible. However, the total observing time amounted to only one night, so that only half of the fields could be observed twice.

During our observing run, the blue CCD (a mosaic composed of two 2k$\times$4k E2V chips) read without any binning, a slit of 1\arcsec\,\footnote{The second night suffered from a bad seeing, which forced us to lengthen the exposure durations, and even use a 1.2\arcsec slit for one of the fields (that centered on Tr16-22). A consequence of this was a change in extraction radius during the reduction (see below).} and the 1200B grating ($R\sim 1400$) were used. We adopted for each field an observing sequence of 8 or 16 subexposures with retarder waveplate positions of $+45^{\circ}$, $+45^{\circ}$, $-45^{\circ}$, $-45^{\circ}$, $+45^{\circ}$, $+45^{\circ}$, $-45^{\circ}$, $-45^{\circ}$,...  Duration of individual subexposures ranges from 10 to 300s, depending on the targets' luminosity and the variable seeing. It was therefore adjusted several times during the run, to remain close, but below, the saturation level (spectropolarimetric investigations requiring the highest possible signal-to-noise ratio). 

%Second, a bad manipulation led to a high saturation of the CCD at the beginning of the third night. As a consequence, one of the two CCDs of the detector mosaic was lost, but fortunately this did not affect the scientific results of our run, since all exposures scheduled that night used only one CCD.

\subsection{Data Reduction}

In recent years, there has been some debate in the literature about the reality of some magnetic field detections claimed using the FORS instruments. To settle the debate, \citet{bag12} has performed an overall reduction of all spectropolarimetric FORS datasets. They notably show the implications of some data reduction choices and we took into account the considerations made in that work. 

Using the observing sequence described above, the normalized Stokes $V/I$ profile can be determined, as well as a diagnostic ``null''  profile \citep{don97,bag09}:\\
\begin{equation}
\begin{array}{rcl}
\pv = V/I &=& {1 \over 2 N} \sum\limits_{j=1}^N \left[ 
\left(\frac{\fo - \fe}{\fo + \fe}\right)_{-45^\circ} - 
\left(\frac{\fo - \fe}{\fo + \fe}\right)_{+45^\circ}\right] \\[2mm]
\nv = N/I &=& {1 \over 2 N} \sum\limits_{j=1}^N (-1)^{(j-1)}\left[ 
\left(\frac{\fo - \fe}{\fo + \fe}\right)_{-45^\circ} - 
\left(\frac{\fo - \fe}{\fo + \fe}\right)_{+45^\circ}\right]\; ,\\
\end{array}
\label{Eq_V_and_N}
\end{equation}
where \fo\ and \fe\ are the counts recorded for the parallel and perpendicular beams, respectively, $I$ is the simple sum of all \fo\ and \fe\ spectra, and $N$ is the number of pairs of subexposures (i.e. sets of $\{-45^\circ$, $+45^\circ \}$ observations). To get the \fo\ and \fe\ for each target, the data were corrected for bias but not flat-fielded, since there is no evidence that flatfielding improves the magnetic field determination \citep{bag12}. Furthermore, the spectra were extracted using a simple aperture extraction since,  in practice, optimal extraction applied to high signal-to-noise ratio leads to noisier spectra \citep[e.g. ][]{bag12}. The aperture extraction radius was fixed to 12\,px for data taken on the first and third night, and between 15 and 24\,px (depending on the actual seeing) for data taken during the second night. Note that sky background was subtracted for each spectrum\footnote{The background subtraction in IRAF (see below) used an average of small sky regions surrounding the star (usually $-35:-25$ and $+25:+35$, or $-55:-45$ and $+45:+55$ when the seeing was bad or a companion existed - only one of these regions being used if the star was too close to a slit edge).}. Finally, wavelength calibration was made by considering arc lamp data taken at only one retarder waveplate position (in our case, $-45^\circ$) for all science frames of a given field, whatever the position of the retarder waveplate actually is for them. As shown by \citet{bag09}, this minimizes spurious polarization signals. Note that, indeed, wavelength calibration remains specific to the actual position of the spectra on the CCD, i.e. it is different from one object to the next as well as for the parallel and perpendicular beams. In all cases, the final, calibrated spectra have 5813\,px with a step of 0.43\AA\ (close to the dispersion value) and a start wavelength of 3500.215\AA, as recommended for this grating.

Whenever it was possible, we reduced our data using two different procedures. For the data obtained in ``fast mode" (i.e. single target), we used both the FORS pipeline\footnote{ftp://ftp.eso.org/pub/dfs/pipelines/fors/fors-pipeline-manual-4.1.pdf} and standard IRAF routines\footnote{$imcombine$ (for master bias creation), $apall$ (for spectrum extraction), and $identify$ (for wavelength calibration, using the wavelength list from the FORS manual which is available at http://www.eso.org/sci/facilities/paranal/instruments/fors/doc/).}. However, the FORS pipeline only accepts datasets with a pair of frames obtained with the retarder waveplate at $+45^{\circ}$ and $-45^{\circ}$, or a quadruplet with retarder waveplate at the position angles $+45^{\circ}$, $-45^{\circ}$, $+135^{\circ}$, and $+225^{\circ}$. Since we adopted a different observing strategy (8 or 16 subexposures taken at two angles, rather than a single pair, see above), we had to combine our 8 or 16 subexposures into 4 or 8 pairs, to reduce these pairs separately, and finally to make use of external procedures to combine the pipeline products into a single result for each 8 or 16 subexposure dataset. Moreover, for polarimetric data obtained in multi-object mode, we could use only IRAF routines, because the FORS pipeline fails to associate the pairs of beams pertaining to individual objects. In the end, the spectra reduced within IRAF appeared very similar to those reduced with the FORS pipeline, when they existed, though the former ones seem slightly less noisy. This is explained by bad seeing, and by the position of the spectra, which, even in ``fast'' mode, were found very close to gaps between slitlets. These problems could be better dealt with through manual fine tuning of extraction and background substraction, than with the pipeline semi-automatic mode. Nevertheless, we decided to use also the FORS pipeline to check our results in two independent ways.
%This can be explained by the bad seeing and position of the spectra (close to gaps, even in ``fast'' mode), which required the use of manual fine tuning for a better extraction. Nevertheless we decided to use also the FORS pipeline to check our results in two independent ways. 

In a final step, the longitudinal magnetic field was estimated by minimising the expression \citep{bag02}:
\begin{equation}
\chi^2 = \sum_i \frac{(y_i - \bz\,x_i - a)^2}{\sigma^2_i}
\label{Eq_ChiSquare}
\end{equation}
where $i$ takes a different value for each spectral point, $a$ is a constant accounting for possible spurious residual polarization in the continuum, $y_i$ is either \pv\ or \nv\ at the wavelength $\lambda_i$, and $x_i = -g_\mathrm{eff}\ \cz \ \lambda^2_i\ 1/I_i\ (\mathrm{d}I/\mathrm{d}\lambda)_i$, with $\cz \simeq 4.67 \times 10^{-13}$\,\AA$^{-1}$\,G$^{-1}$ and $g_\mathrm{eff}$ the effective Land\'{e} factor, assumed to be 1 near the Balmer lines of hydrogen and 1.2 elsewhere. 

In this equation, $(\mathrm{d}I/\mathrm{d}\lambda)_i$ was evaluated through $(I_{i+1}-I_{i-1})/(\lambda_{i+1}-\lambda_{i-1})$, as by \citet{bag02}, while the error on \pv\ is estimated through $\sigma_i=\frac{1}{2\sqrt{N}} \, \frac{\sigma(f)_i}{f_i}$ \citep[see eq. A6 of][]{bag09}. Following the FORS pipeline manual\footnote{ftp://ftp.eso.org/pub/dfs/pipelines/fors/fors-pipeline-manual-4.1.pdf}, the error on each spectrum (\fo\ or \fe) amounts to $\sigma(f)_i=\sqrt{f_i/g}$ where $f_i$ is the flux at the wavelength $\lambda_i$ and $g$ is the gain in electrons per ADU (0.53 or 0.58, depending on the CCD chip) - the read-out noise is neglected because of its small value compared to the high level of the stellar flux. Note that the FORS pipeline accounts for the error introduced by background subtraction and by the extraction, whereas for data reduced with IRAF routines we considered only the contribution due to photon-noise.
%In principle, the FORS pipeline should also account for the error introduced by background subtraction and by the extraction.
%The error on $I$ is then easily deduced assuming error propagation (since $I(\lambda_i)=\sum f_j(\lambda_i)$, $\sigma(I)_i=\sqrt (\sum \sigma(f)_i)$).

A few additional remarks must be made at this stage. First, we have estimated the magnetic field for the ``raw'' \pv\ and \nv\ profiles, and for the same \pv\ and \nv\ profiles rectified using a Fourier filter of width 300px \citep{bag12}. That correction results only in a minimal difference (about 15G on average), well within the error bars (typically 80G), confirming the results of \citet[see their Table 2]{bag12}. Second, some values of the \pv\ and \nv\ profiles were discarded: the 3\% bluest and 3\% reddest of the wavelength range with recorded data, as well as the spectral bins where \nv\ deviates more than 3$\sigma$ from zero (and the two adjacent bins), as in \citet{bag12}. In addition, we did not consider the spectral bins with $x_i$ outside the range $\{-1\times 10^{-6}$, $+1\times 10^{-6} \}$, as they were small in number and erratic in value but dominating the slope determination. Third, we derived magnetic field values for the full spectral range and for selected spectral windows (avoiding continuum regions, or regions affected by ISM/nebular features, see online material in Appendix). Again, this results in small changes (about 45G on average), within the error bars. Finally, \citet[and references therein]{bag12} found several hints that the error bars of the magnetic field determinations made using the FORS instrument were underestimated. Therefore, the error on \bz, derived as is usual from the linear interpolation, was corrected following the method of \citet{bag12}, to take into account the actual scattering around the fitted line. This correcting factor is $\sqrt{\chi^2_{\rm min}/\nu}$, where $\chi^2_{\rm min}$ is the minimum value of the $\chi^2$ derived as in Eq.~(\ref{Eq_ChiSquare}) and $\nu$ is the number of degrees of freedom of the system (i.e., the number of spectral points minus two).

\section{Results}

The derived values of the longitudinal fields for all targets are listed in Table \ref{result}, which also gives for comparison the \nz\ values found from the null profiles. For comparison purposes, values with or without rectification, for the full spectral range or considering only specific spectral windows are given. Spectral types and X-ray luminosity, reproduced from \citet{naz11}, are also shown. Several conclusions can now be drawn.

First, even though the FORS pipeline results are about 10-15\% noisier then those obtained with manual data reduction, all results are consistent with each other (i.e. the conclusion of a detection or a non-detection is found with both methods), and the field values are within 2$\sigma$ error bars of each other. Second, the diagnostic \bz\ derived from the null profile are consistent with zero (at the 3$\sigma$ level in the worst cases, most often at the 1$\sigma$ level), as is expected, confirming the quality of our data reduction. Finally, and most importantly, two stars shows a magnetic field detection at the 6$\sigma$ level and with a strength $>$300\,G, two features suggesting a true detection \citep{bag12}. The other 19 stars have \bz\ compatible with zero at the 3$\sigma$ level, as do the diagnostic profiles. 

\begin{figure*}
\includegraphics[width=8cm]{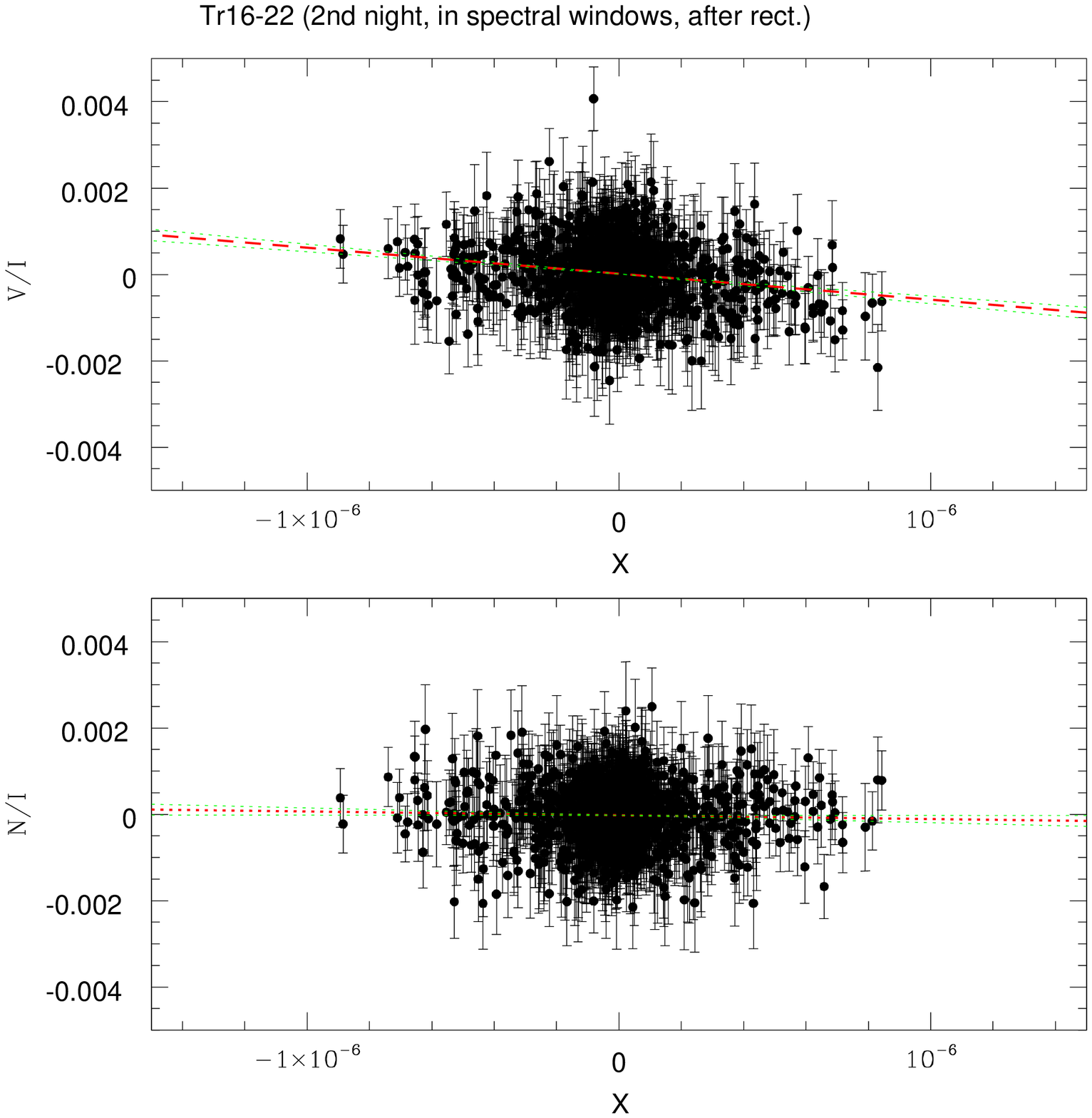}
\includegraphics[width=8cm]{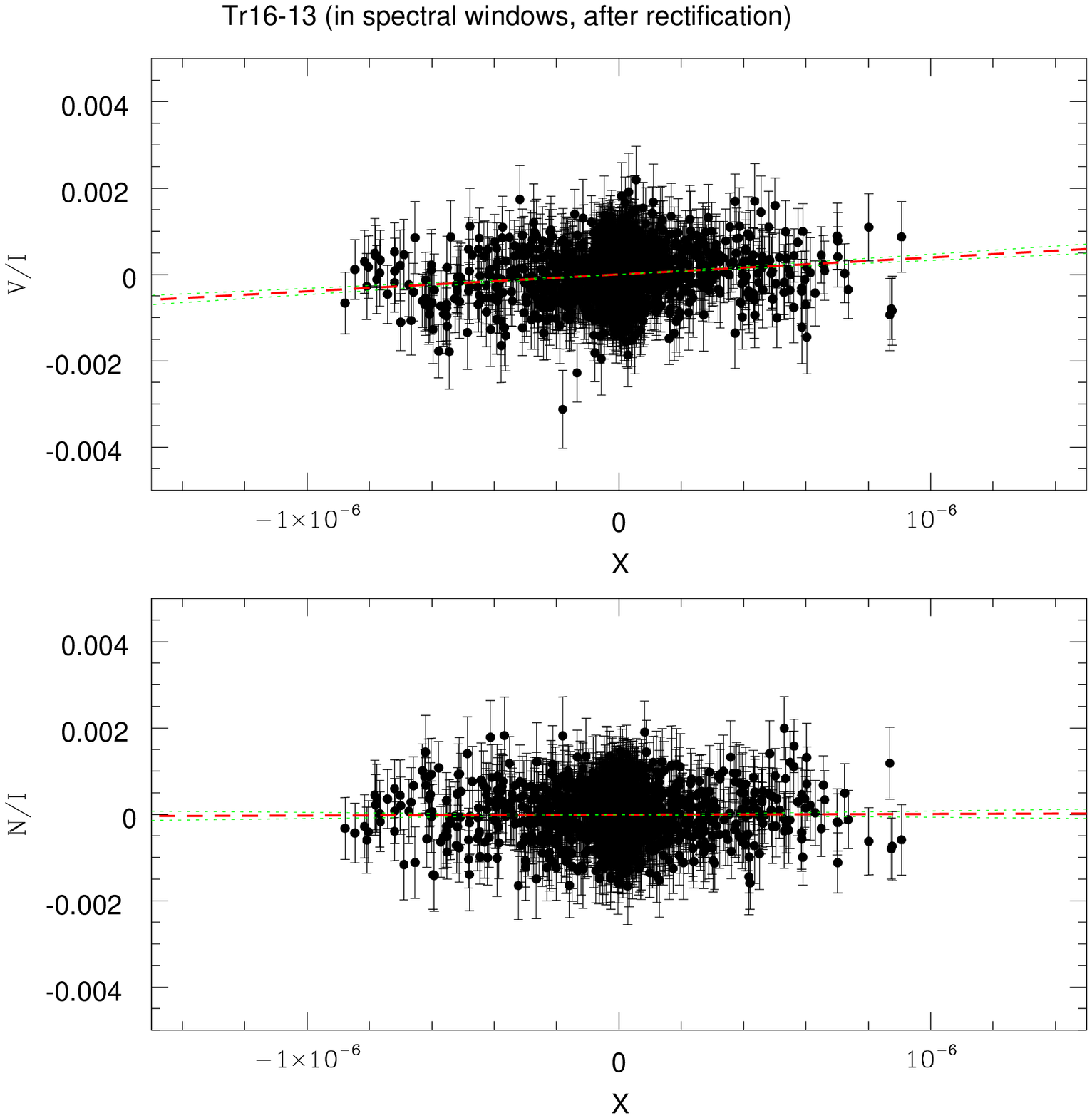}
\caption{Normalized Stokes V (in selected spectral windows, with rectification) as a function of x ($= -g_\mathrm{eff}\ \cz \ \lambda^2\ 1/I\ (\mathrm{d}I/\mathrm{d}\lambda)$, see Eq.~(\ref{Eq_ChiSquare})) for the two positive cases of Tr16-22 (left) and Tr16-13 (right). In these panels, the slope is directly linked to the intensity of the longitudinal field. The best-fit lines, derived from the least-squares analysis, are drawn with thick dashed lines, for both the actual Stokes V parameter (top panels) and the diagnostic null profile (bottom panels). In the latter case, or more generally in the case of non-detections, these lines should be horizontal.  The line thickness is larger than the 1$\sigma$ error bar on the constant term, while the two thin dotted lines show lines with slopes at $\pm 1 \sigma$ from the best-fit value.}
\label{stokesv}
\end{figure*}

\begin{table*}
\begin{center}
\caption{Field measurements for all stars of our sample, for both Stokes $V$ profiles and their related diagnostic null profiles. For fields with only one target, the results found from the FORS pipeline are also quoted at the bottom of the table. A star after the name indicates the main targets, those displaying X-ray peculiarities, and boldfaces shows the 6$\sigma$ detections.}
\label{result}
\begin{tabular}{l l c | r r | r r | r r }
            \hline\hline
Target & Sp. Type & \loglxlb & \multicolumn{2}{|r|}{Full range, non-rect.} & \multicolumn{2}{r|}{Spectral windows, non-rect.} & \multicolumn{2}{r}{Spectral windows, rectified}\\
 & & & \bz & \nz & \bz & \nz & \bz & \nz \\
 & & & (G) & (G) & (G) & (G) & (G) & (G) \\
\hline
\multicolumn{9}{c}{IRAF} \\
Coll228-66           & O9.5V   & $-7.5$  & $-88\pm43$ & $14\pm42$  & $-72\pm45$ & $-6\pm44$  & $-63\pm45$  & $-6\pm44$  \\
Coll228-68$\star$    & B1Vn    & $-6.5$  & $31\pm56$  & $-31\pm51$ & $60\pm58$  & $-48\pm55$ & $78\pm58$   & $-46\pm55$ \\
HD\,93097            & B0Vn    & $<-7.9$ & $-57\pm44$ & $-3\pm44$  & $-89\pm47$ & $31\pm48$  & $-80\pm46$  & $1\pm47$   \\
HD\,93190$\star$     & B0IV:ep & $-7.8$  &$-120\pm93$ & $-10\pm90$ &$-154\pm120$& $-25\pm120$& $-199\pm128$& $26\pm125$ \\
HD\,93250-n1$\star$  &O4III(fc)& $-6.4$  & $96\pm78$  &$123\pm75$  & $90\pm85$  & $94\pm83$  & $58\pm82$   & $96\pm82$  \\
HD\,93250-n2$\star$  &         &         & $25\pm69$  & $49\pm69$  & $17\pm74$  & $71\pm72$  &  $8\pm73$   & $43\pm71$  \\
HD\,93501-n1$\star$  & B1.5III:& $-6.9$  &  $9\pm74$  & $201\pm72$ & $-3\pm81$  & $267\pm79$ & $-3\pm81$   & $272\pm78$ \\
HD\,93501-n2$\star$  &         &         & $60\pm70$  & $-4\pm69$  & $12\pm76$  & $-58\pm75$ & $18\pm76$   & $-66\pm74$ \\
HD\,305524-n1        & O7V((f))& $-7.3$  & $-50\pm71$ & $-96\pm69$ & $-65\pm79$ & $-72\pm75$ & $-16\pm76$  & $-79\pm75$ \\
HD\,305524-n2        &         &         & $-120\pm57$& $-13\pm51$ & $-97\pm68$ & $-3\pm58$  & $-55\pm64$  & $1\pm58$   \\
HD\,305534-n1        & B0.5V+B1V:& $-7.9$& $-4\pm78$  & $-124\pm78$& $14\pm87$  & $-197\pm87$& $37\pm86$   & $-199\pm87$\\
HD\,305534-n2        &           &       & $-19\pm59$ & $-39\pm56$ & $-46\pm66$ & $-86\pm65$ & $-63\pm65$  & $-88\pm65$ \\
LS\,1853-n1          & B1Ib    & $<-9.0$ & $41\pm60$  & $12\pm60$  & $48\pm64$  & $-6\pm62$  & $48\pm63$   & $-9\pm62$  \\
LS\,1853-n2          &         &         & $-6\pm52$  & $-103\pm51$& $18\pm55$  & $-116\pm55$& $24\pm54$   & $-118\pm55$\\
LS\,1865-n1$\star$   &O8.5V((f))& $-6.9$ & $-72\pm67$ & $-83\pm67$ & $-41\pm72$ & $-84\pm69$ & $-56\pm72$  & $-86\pm70$ \\
LS\,1865-n2$\star$   &          &        & $-70\pm52$ & $64\pm51$  & $-98\pm56$ & $92\pm52$  & $-112\pm56$ & $95\pm52$  \\
Tr16-2               & B1V     & $<-8.0$ & $120\pm62$ & $33\pm61$  & $82\pm66$  & $62\pm65$  & $55\pm65$   & $51\pm65$  \\
Tr16-4               & B1V     & $<-7.9$ & $1\pm59$   & $113\pm60$ & $29\pm63$  & $100\pm62$ & $21\pm63$   & $101\pm62$ \\
Tr16-5$\star$        & B1V     & $-6.4$  & $86\pm76$  & $-115\pm74$& $42\pm85$  & $-131\pm83$& $19\pm85$   & $-128\pm83$\\
Tr16-10$\star$       & B0V     & $-7.0$  & $124\pm102$& $60\pm100$ & $121\pm108$& $116\pm108$& $153\pm108$ & $114\pm108$\\
Tr16-11$\star$       & B1.5V   & $-6.4$  &$-104\pm201$&$-177\pm205$&$-155\pm225$& $168\pm230$& $-145\pm229$& $175\pm230$\\
Tr16-13              & B1V     & $-7.8$  & $\bld{490\pm70}$ & $-7\pm64$  & $\bld{456\pm76}$ & $13\pm71$  & $\bld{393\pm73}$  & $16\pm71$  \\
Tr16-14              & B0.5V   & $-8.0$  &$-143\pm166$& $161\pm165$& $91\pm186$ & $433\pm191$& $120\pm187$ & $418\pm190$\\
Tr16-15              & B0V     & $-7.9$  & $18\pm68$  & $24\pm68$  & $-1\pm73$  & $65\pm73$  & $-6\pm73$   & $67\pm72$  \\
Tr16-22-n1$\star$    & O8.5V   & $-6.3$  & $\bld{-386\pm70}$& $3\pm68$   &$\bld{-433\pm73}$ & $54\pm73$  & $\bld{-454\pm72}$ & $45\pm74$  \\
Tr16-22-n2$\star$    &         &         & $\bld{-450\pm81}$& $-28\pm76$ &$\bld{-578\pm88}$ & $-94\pm83$ & $\bld{-604\pm87}$ & $-88\pm82$ \\
Tr16-64$\star$       & B1.5Vb  & $-6.1$  & $-43\pm55$ & $17\pm53$  & $-49\pm58$ & $41\pm57$  & $-65\pm57$  & $43\pm57$  \\
Tr16-115             & O9.5V   & $-7.2$  & $-15\pm67$ & $38\pm66$  & $-31\pm69$ & $23\pm67$  & $-38\pm68$  & $22\pm67$  \\
\hline
\multicolumn{9}{c}{FORS pipeline} \\
HD\,93190$\star$     & B0IV:ep & $-7.8$  & $-272\pm110$ & $-23\pm104$ & $-192\pm141$&$-45\pm138$& $-154\pm140$& $2\pm140$  \\
HD\,93250-n1$\star$  &O4III(fc)& $-6.4$  & $102\pm91$   & $120\pm87$  & $95\pm98$   & $92\pm96$ & $68\pm95$   & $78\pm95$  \\
HD\,93250-n2$\star$  &         &         & $-128\pm137$ & $82\pm116$  & $-250\pm144$& $93\pm126$& $-309\pm139$& $47\pm122$ \\
HD\,93501-n1$\star$  & B1.5III:& $-6.9$  & $13\pm85$    & $200\pm81$  & $-3\pm93$   & $295\pm90$& $22\pm92$   & $309\pm89$ \\
HD\,93501-n2$\star$  &         &         & $90\pm84$    & $41\pm78$   & $77\pm93$   & $3\pm87$  & $84\pm91$   & $-1\pm86$  \\
Tr16-22-n1$\star$    & O8.5V   & $-6.3$  & $\bld{-463\pm83}$  & $-43\pm76$  & $\bld{-575\pm91}$ & $3\pm81$  & $\bld{-620\pm89}$ & $-5\pm82$  \\
Tr16-22-n2$\star$    &         &         & $\bld{-483\pm92}$  & $29\pm86$   & $\bld{-604\pm101}$& $0\pm97$  & $\bld{-632\pm99}$ & $-8\pm98$  \\
\hline
\end{tabular}
\end{center}
{\footnotesize {1. Comparison of names in Simbad or \citet{naz11} and those of this paper: CPD -59255B = Coll228-66, Cl* Trumpler 14 MJ 449 = LS\,1865, Cl* Trumpler 14 MJ 366 = LS\,1853, Cl* Trumpler 16 MJ 506 = Tr16-2, Cl* Trumpler 16 MJ 466 = Tr16-4, Cl* Trumpler 16 MJ 427 = Tr16-5, Cl* Trumpler 16 MJ 327 = Tr16-10, Cl* Trumpler 16 MJ 289 = Tr16-11, Cl* Trumpler 16 MJ 339 = Tr16-13, Cl* Trumpler 16 MJ 357 = Tr16-14, Cl* Trumpler 16 MJ 372 = Tr16-15, Cl* Trumpler 16 MJ 496 = Tr16-22, Cl* Trumpler 16 MJ 554 = Tr16-115.\\ 
2. Note that the typical \loglxlb\ in Carina Chandra data is $-7.2$ \citep{naz11}.}}
\end{table*}

\subsection{Tr16-22}
With the Chandra and XMM observations of the Carina nebula, Tr16-22 has been singled out as a peculiar object. Indeed, it is strongly overluminous in X-rays \citep[and references therein]{eva04,ant08,naz11}. Moreover, this high-energy emission is unusually hard for an O-type star, with a component at 2\,keV of a strength similar to that of the soft component at 0.7\,keV \citep{naz11}. Furthermore, the X-ray luminosity varies by a factor of 30 \citep{com11}, whereas the X-ray emission of O-stars is known to be rather stable.

\begin{figure}
\includegraphics[width=8cm]{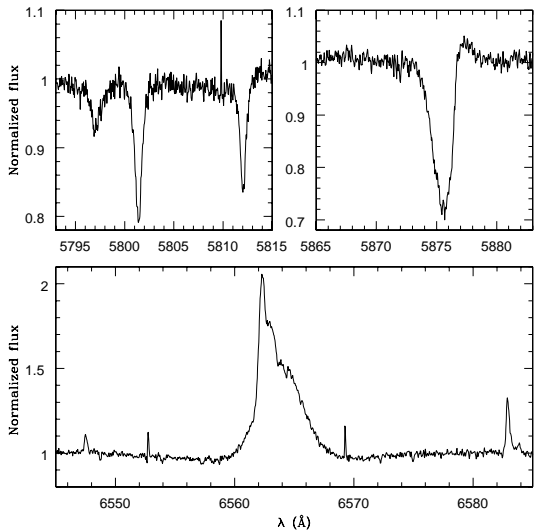}
\caption{Archival FEROS spectrum of Tr16-22, showing the narrow metallic lines, the P cygni profile of He\,{\sc i}\,5876\AA, and the strong \ha\ emission. Note that the narrowest emission component in \ha\ is nebular in origin (other nebular lines, e.g. those of [N\,{\sc ii}], are indeed detected). }
\label{feros}
\end{figure}

Radial velocity variations have been reported for this object from low-resolution monitorings \citep{com11,wil11}, suggesting the star to be a binary. With the information at hand before this study, X-rays linked to a wind-wind collision might at first seem a good explanation for the observed high-energy characteristics. However, the observed large overluminosity ($>$1\,dex) is neither observed nor theoretically expected in late-type O+OB colliding-wind binaries \citep[for examples of O+O binary X-ray observations, see][]{naz09,naz11}. Moreover, the reported spectral changes, observed with low signal-to-noise \citep{com11}, could equally well be compatible with line profile variations. In view of these facts and our spectropolarimetric result (Table \ref{result} and Fig. \ref{stokesv}), magnetically-confined winds now appear as a much better explanation \citep{eva04,naz11}. A monitoring is now needed to further constrain the stellar, wind, and magnetospheric properties which would enable us to perform a full modelling of this star. In this context, we have examined the single archival FEROS spectrum of Tr16-22. This spectrum appears quite similar to that of the sharp-lined magnetic O9.5IV star HD\,57682 \citep{gru09}, though the lines of Tr16-22 seem even narrower than those of HD\,57682. A long period for Tr16-22 would not be surprising, as an intense field is able to brake the rotation of massive stars with strong stellar winds \citep{udd09}. Indeed, many magnetic O stars are slow rotators (see e.g. the extreme cases of 538d for HD\,191612 or even 55yrs for HD\,108, \citealt{naz10}). The FEROS spectrum also reveals a strong emission component in \ha\ and a P Cygni profile for He\,{\sc i}\,5876\AA\ (Fig. \ref{feros}), two features comparable to those of the strongly confined $\theta^1$\,Ori\,C and of the Of?p stars. This reinforces our suspicion that Tr16-22 shows several evidence for the presence of magnetically confined winds.

\subsection{Tr16-13}
This object displays a clear magnetic detection (Table \ref{result} and Fig. \ref{stokesv}), but it could only be observed once, so that we have no independent confirmation of our result. Nevertheless, the obtained field values clearly differ between the actual Stokes V profile and the null profile, leaving little doubt on the detection. Spectra of targets observed at the edge of the field of view may be slightly depolarized \citep{bag02}, but this problem would decrease the absolute field value, rather than causing a spurious detection, thus we are confident in our result. 
%Magnetic measurements of off-axis targets may be affected by depolarization, but the {\it detection} capability is not impacted \citep{bag02} and we are thus confident in our result.
%The actual value, however, may not be exactly that listed in Table \ref{result}: as explained by \citet{bag02}, the FORS instrument is especially suited to detect magnetic fields, but has difficulties in ascertaining their values, especially for off-axis sources, when the multi-object mode is used, which is the case of Tr16-13. 

Tr16-13 is not known to exhibit any peculiarity (e.g. spectroscopic, photometric, or radial velocity variations), but the literature concerning this star is very sparse (e.g. no high-resolution spectra appear to exist). The FORS spectrum does not show pronounced He chemical peculiarities, but this is not unusual amongst early B-type magnetic stars (see e.g. HD\,66665, \citealt{pet11}). The spectral lines of Tr16-13 appear somewhat broad (FWHM of about 170\,\kms), but at the limit of the FORS resolution: convolved to the same resolution, the spectrum of $\tau$\,Sco, a sharp-lined, magnetic B0.5V star, appears similar but with narrower lines. However, contrary to that object, Tr16-13 emits few X-rays. Line profile variations may be expected for this star, and follow-up investigations are thus clearly needed to pinpoint its physical properties.

\subsection{HD\,93250}
The non-detection of a magnetic field may naively appear particularly puzzling in the case of HD\,93250. This object indeed displays non-thermal radio emission \citep{lei95}. Such radiation is known to arise from relativistic electrons accelerated at the shocks between the two winds in a binary system \citep{van}. Moreover, HD\,93250 displays large X-ray variations \citep{rau09}, which reinforces the suspicion of a binary nature. Though radial velocity variations have never been observed up to now \citep[see][and references therein]{rau09,wil11}, a close companion was recently detected using interferometry \citep{san11}. HD\,93250 now appears composed of two similar stars, whose winds collide, in a long-period, eccentric orbit. A magnetic field is indeed required in this system, to explain the synchrotron radio emission of the system. However, the nature and origin of the field - whether associated directly to the star(s) or produced within the shock - is unclear. Considering the first case, it is true that our observations do not detect any organized field at the photosphere, but it must be noted that a low inclination of the star's magnetic axis or its binary nature could impact on our detection capabilities. Moreover, the derived 3$\sigma$ limit on the longitudinal field of HD\,93250 is about 300G, and no theoretical expectation has been published to which we could compare this value. Depending on the exact stellar and wind properties (which are not known with precision), as well as local characteristics of the wind collision zone (equally unknown), which may produce a field amplification \citep{bel04,gon12}, it may well be that our non-detection of an organized magnetic field is not at odds with the presence of the $\sim$1\,G field needed at the collision zone to generate some non-thermal radio emission. 
%Definitely, a better constrain of the system properties is needed before one can claim that our \bz\ limit contradicts with the production of synchrotron emission at the wind-wind shock.

\section{Summary and Conclusion}
Following the detection of hard and bright X-rays in a handful of early-type stars in the Carina nebula, we performed a small spectropolarimetric survey of 21 massive stars in the region using the FORS instrument. Previously, only 7 OB stars in Carina (HD\,93128, HD\,93129B, HD\,93204, HD\,93403, HD\,93843, HD\,303308, HD\,303311) had been examined by \citet{hub11}, alas with no clear detection and only one 3$\sigma$ detection \citep[which is a low detection level for FORS data, see ][]{bag12}.

With this paper, we have quadrupled the number of Carina massive stars searched for magnetic fields. For 19 objects, we could put limits of a few hundreds gauss on magnetic fields, but two objects led to a clear (6$\sigma$) detection: Tr16-22 ($\bz \sim 500$\,G) and Tr16-13 ($\bz \sim 450$\,G). 

Though the late-O star Tr16-22 could have a companion, its hard, variable X-ray emission appears too bright to originate in a wind-wind collision inside the binary. The magnetic field detection (confirmed by two independent observations) rather supports a scenario where magnetically-confined winds collide at the magnetic equator, thereby producing copious hard X-ray emission. An independent confirmation comes from its strong \ha\ emission, similar to that observed in other magnetic cases (the Of?p stars and $\theta^1$\,Ori\,C). With its peculiar X-ray properties, Tr16-22 may thus be seen as an analog to the famous $\theta^1$\,Ori\,C, the best example amongst magnetic O-stars for a good match with theoretical expectations, especially in the X-ray range \citep{gag05}: as a potential second case, Tr16-22 appears as a key in our understanding of confined winds. A full monitoring of this star (X-ray, spectroscopy, spectropolarimetry,...) is thus urgently needed.

Tr16-13 is a poorly known star with no known peculiarities and rather faint X-ray emission. Certainly, more investigation is now needed to ascertain the impact of magnetic fields on its evolution and on its stellar wind. Again, a dedicated monitoring is required to determine the physical parameters of the star as well as the magnetospheric properties, needed before any modelling can take place.

\section*{Acknowledgments}
The authors acknowledge the referee, I.D. Howarth, for his useful comments and promptness to review the paper. We also thank A. David-Uraz, C. Folsom, M. Shultz for useful comments and contributions for a follow-up ESO proposal.
YN acknowledges support from the Fonds National de la Recherche Scientifique (Belgium), the Communaut\'e Fran\c caise de Belgique, the PRODEX XMM and Integral contracts, and the `Action de Recherche Concert\'ee' (CFWB-Acad\'emie Wallonie Europe). VP acknowledges support from the Fonds Qu\'ebecois de la Recherche sur la Nature et les Technologies. GAW ackowledges Discovery grant support from the Natural Science and Engineering Research Council of Canada. ADS and CDS were used for preparing this document.

\appendix

\section[]{Online material}
The following figures show the spectra of the 21 objects, with the spectral windows used for the field determinations (see Sect. 2.2). 

\begin{figure*}
\includegraphics[width=8cm, bb=60 430 565 720, clip]{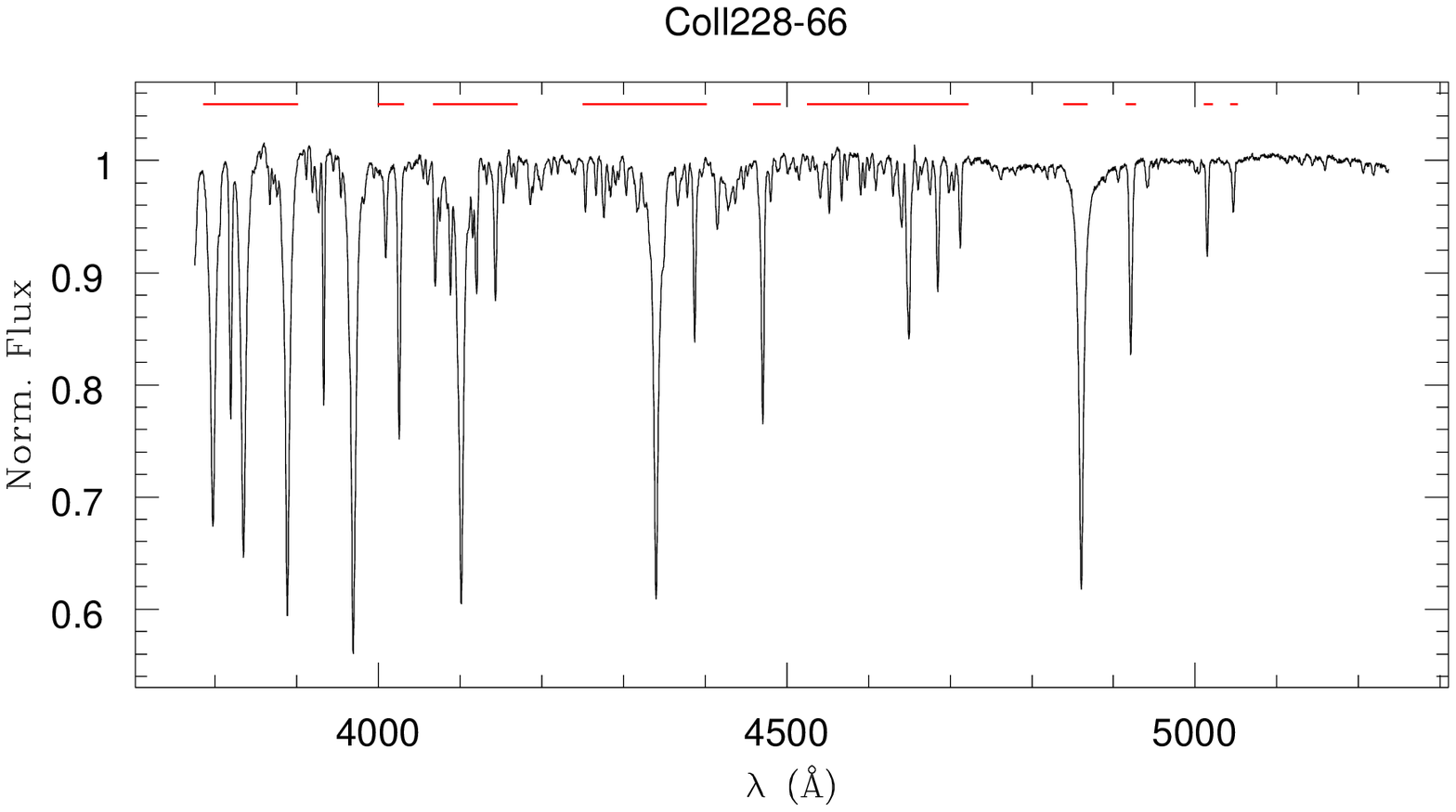}
\includegraphics[width=8cm, bb=60 430 565 720, clip]{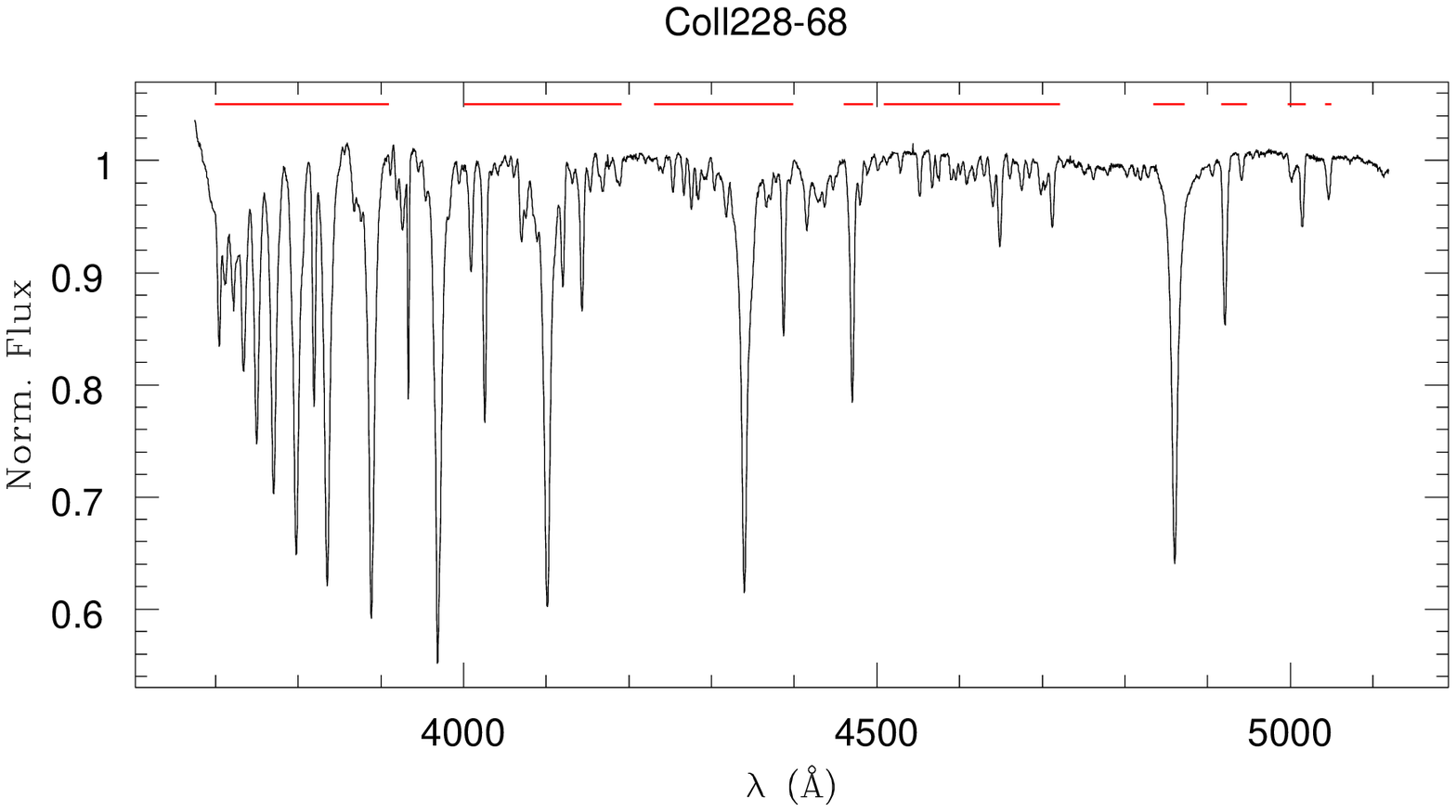}
\includegraphics[width=8cm, bb=60 430 565 720, clip]{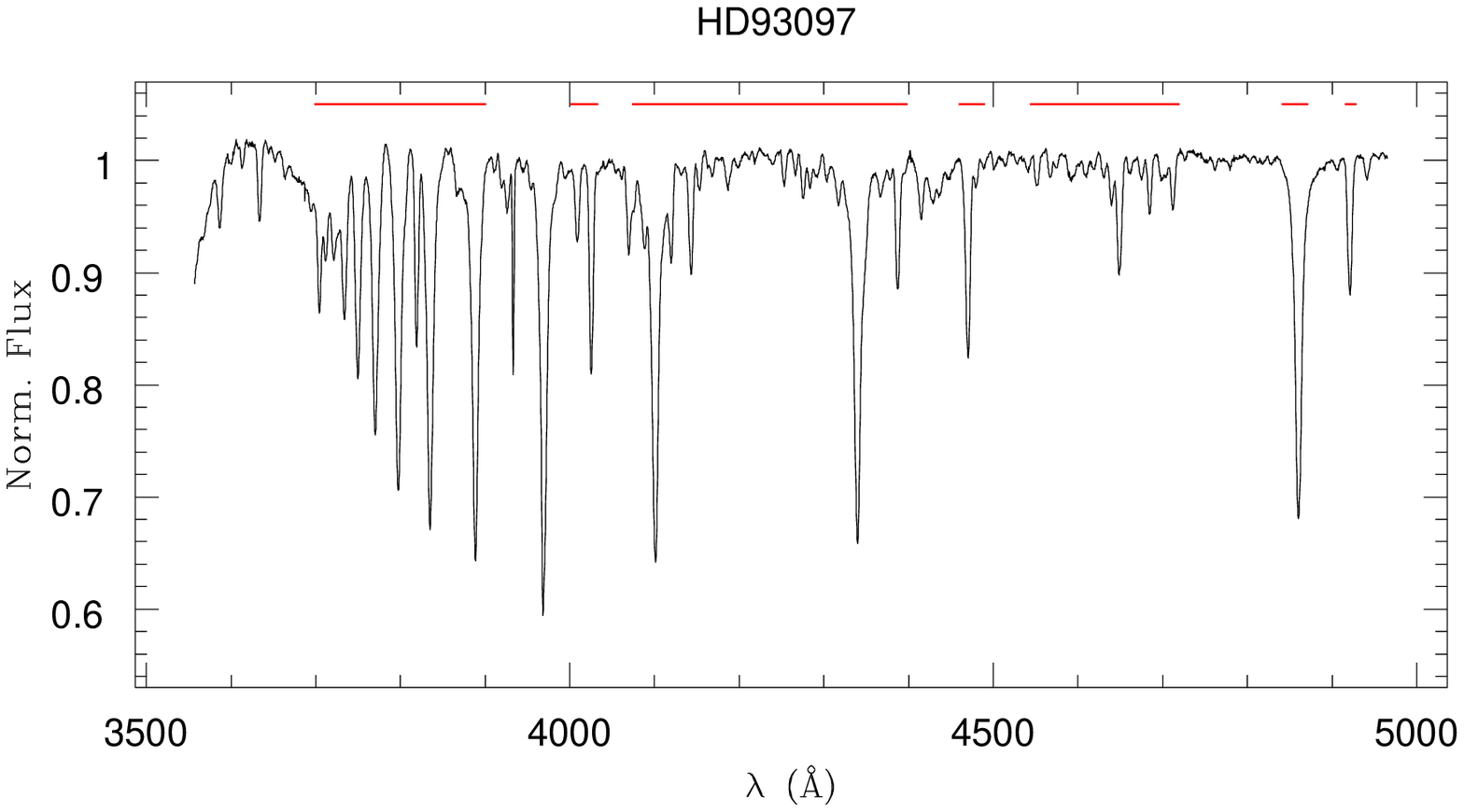}
\includegraphics[width=8cm, bb=60 430 565 720, clip]{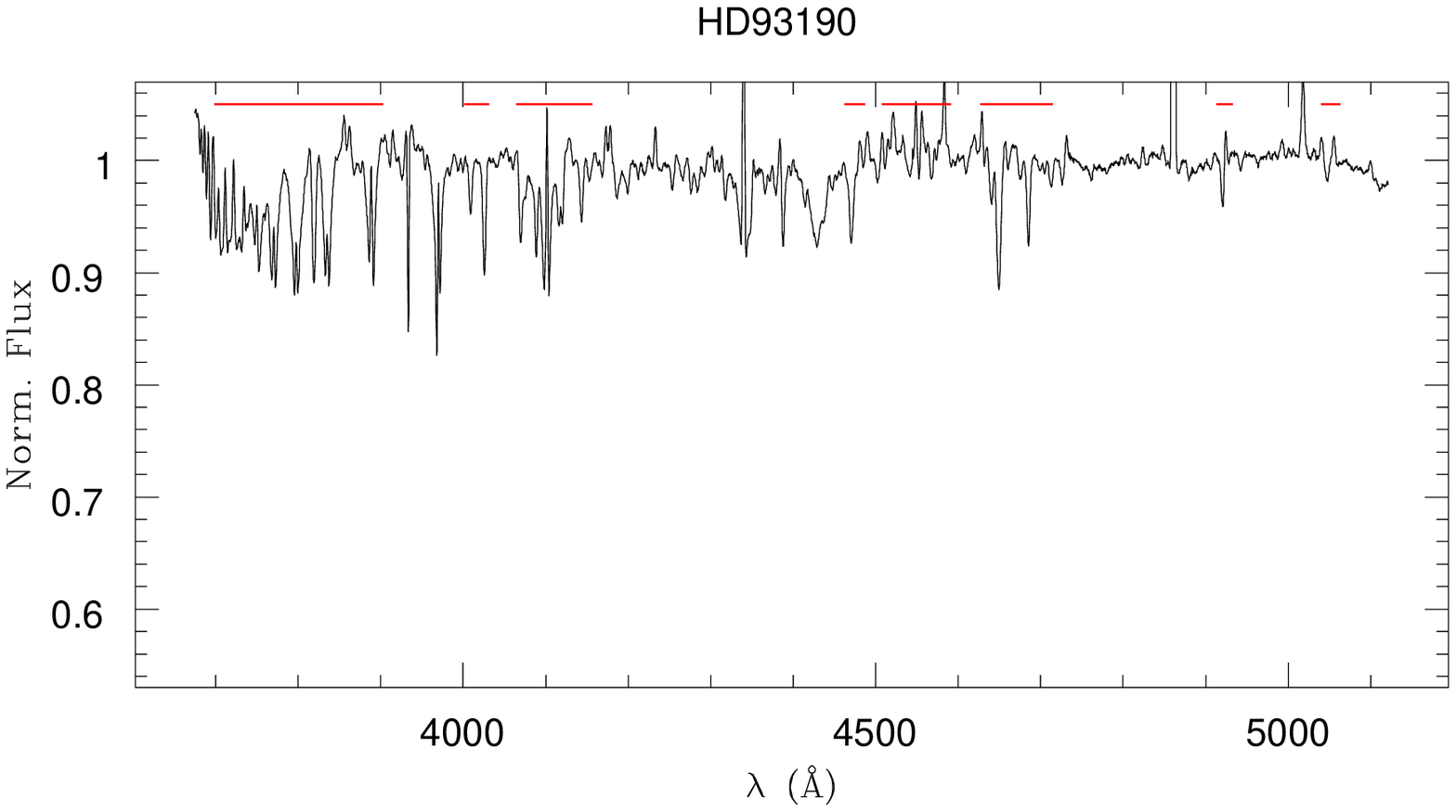}
\includegraphics[width=8cm, bb=60 430 565 720, clip]{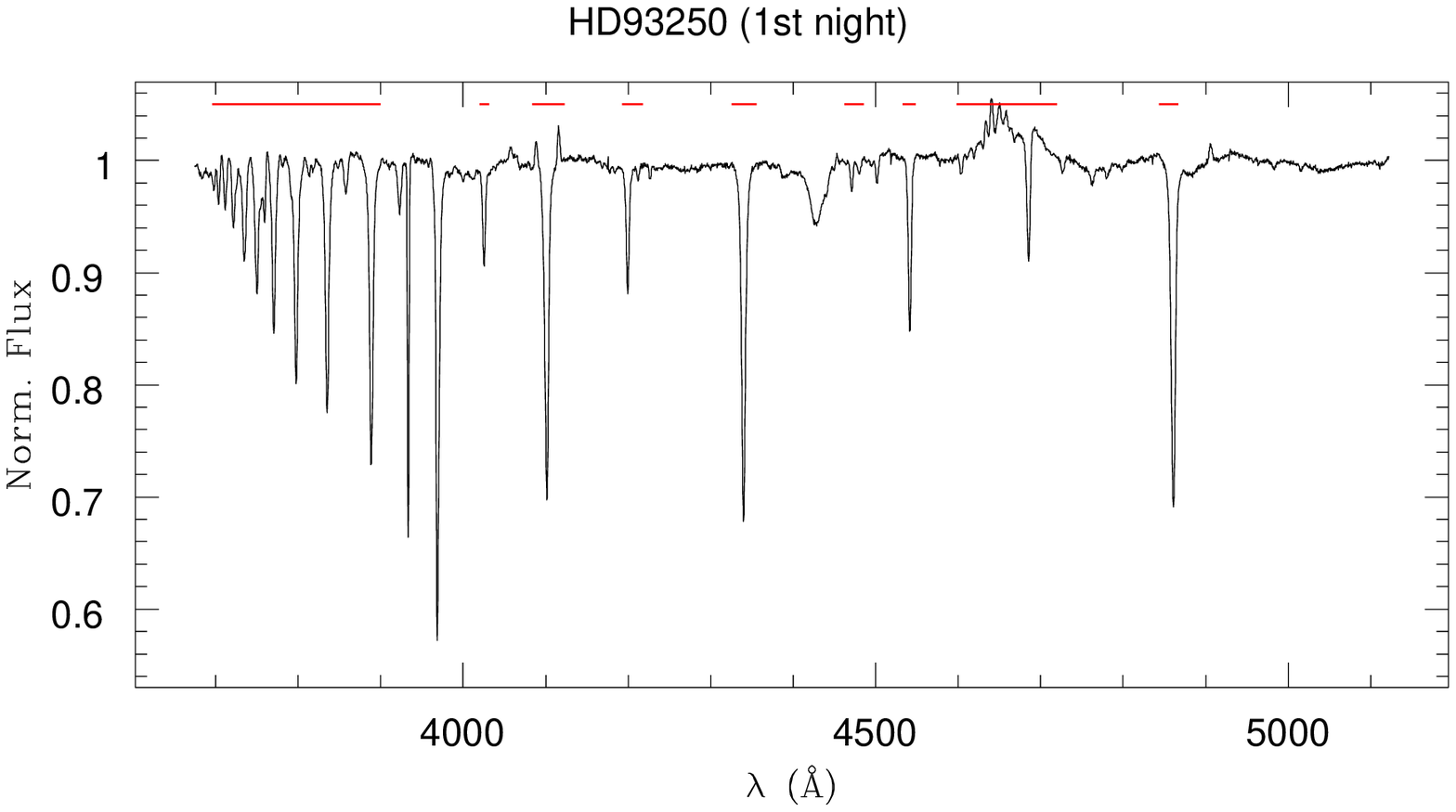}
\includegraphics[width=8cm, bb=60 430 565 720, clip]{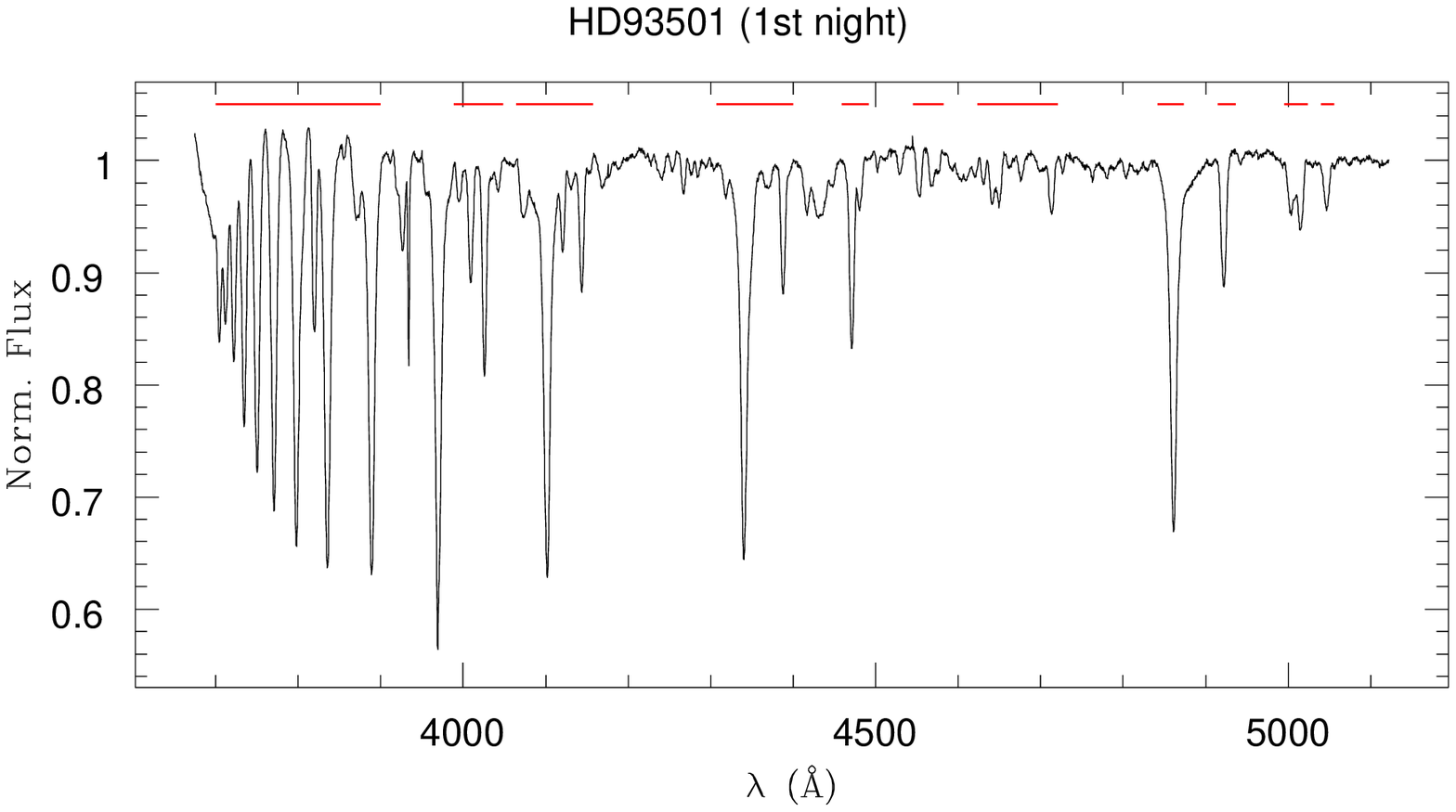}
\includegraphics[width=8cm, bb=60 430 565 720, clip]{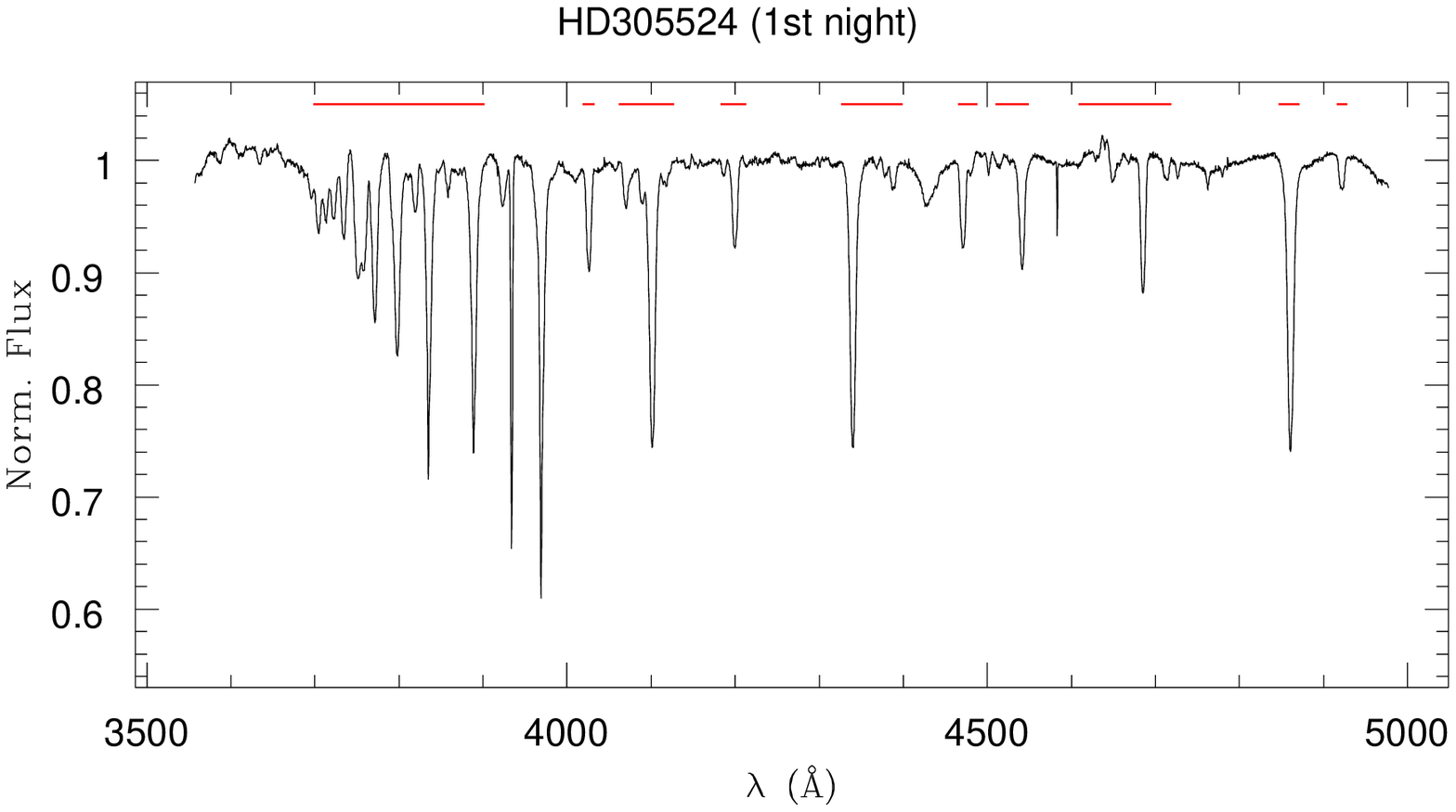}
\includegraphics[width=8cm, bb=60 430 565 720, clip]{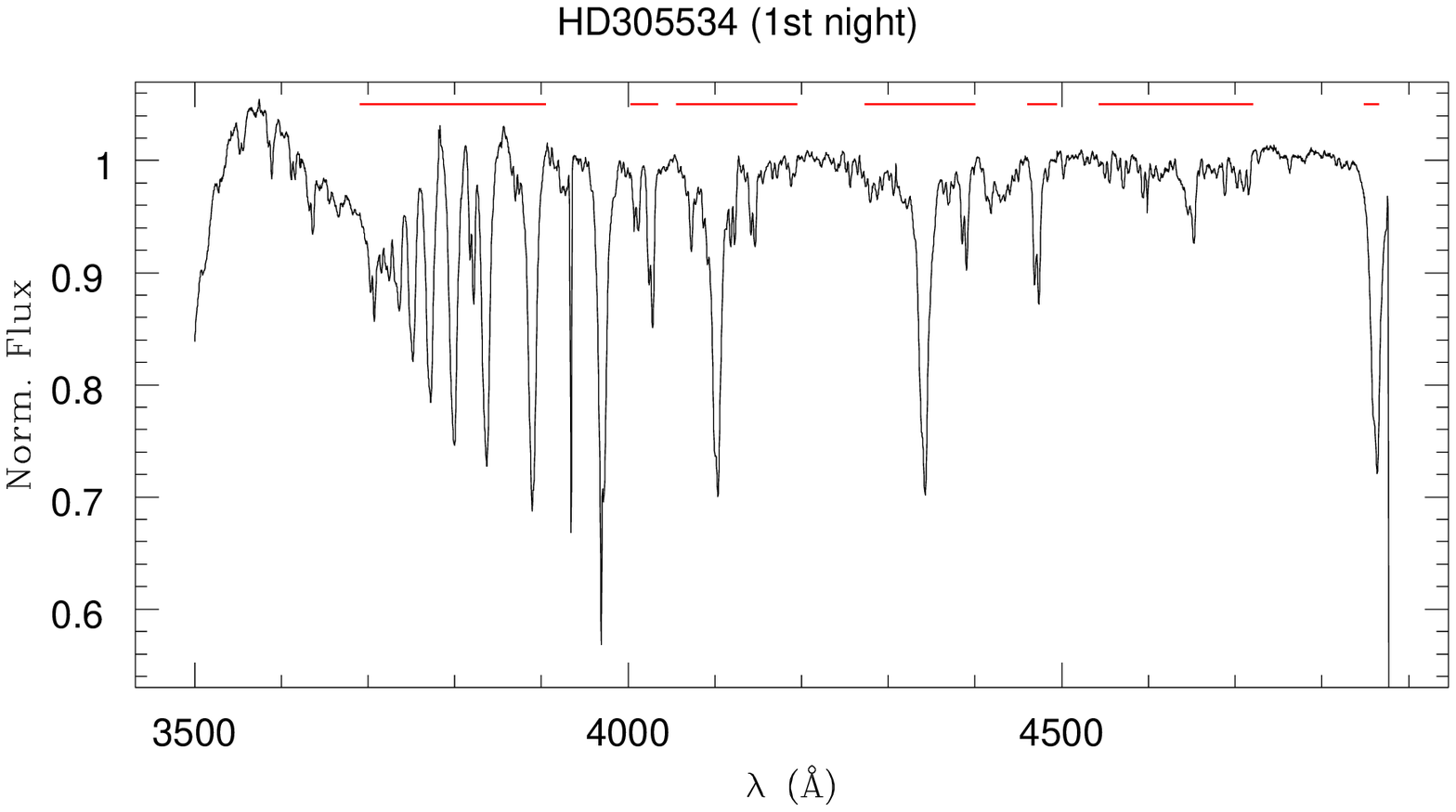}
\includegraphics[width=8cm, bb=60 430 565 720, clip]{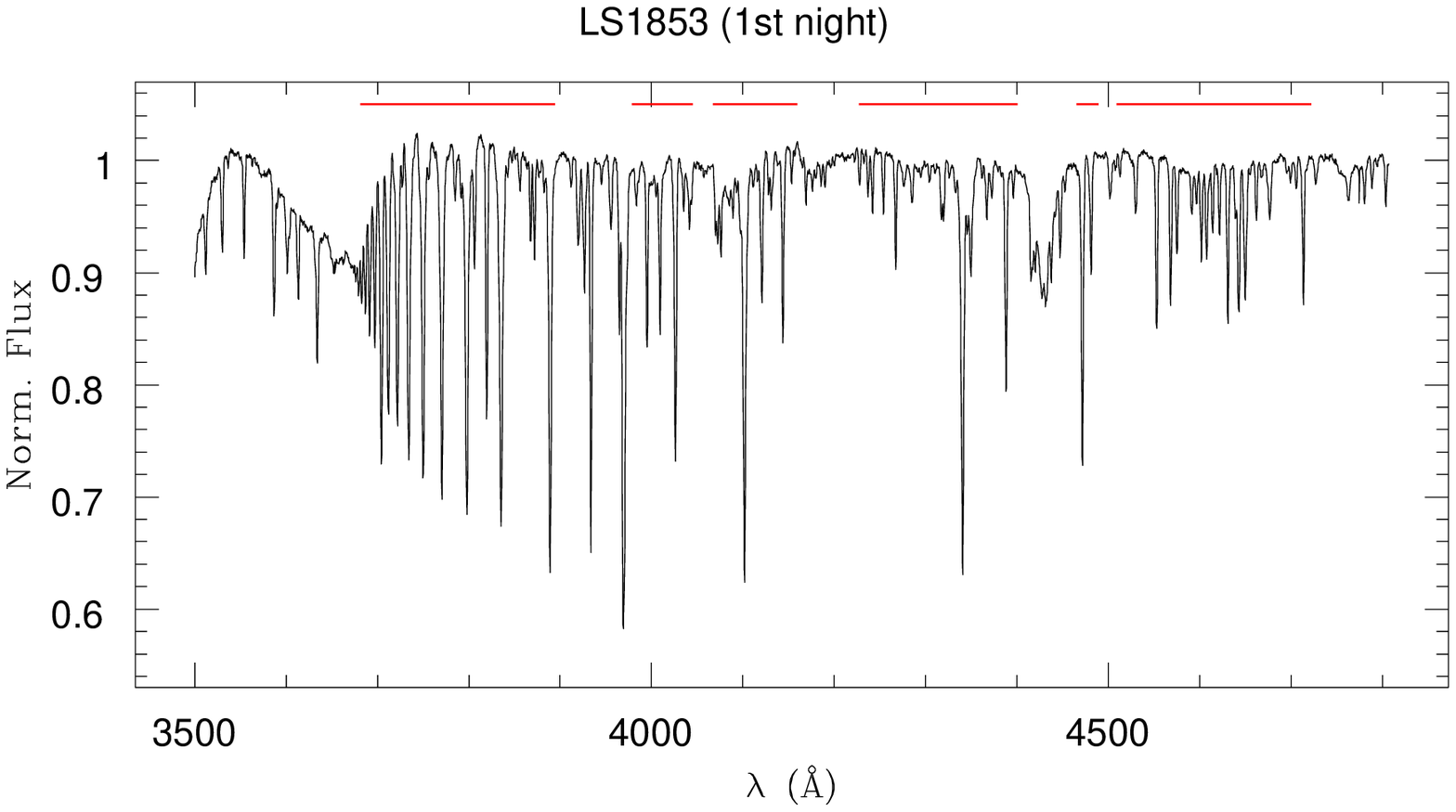}
\includegraphics[width=8cm, bb=60 430 565 720, clip]{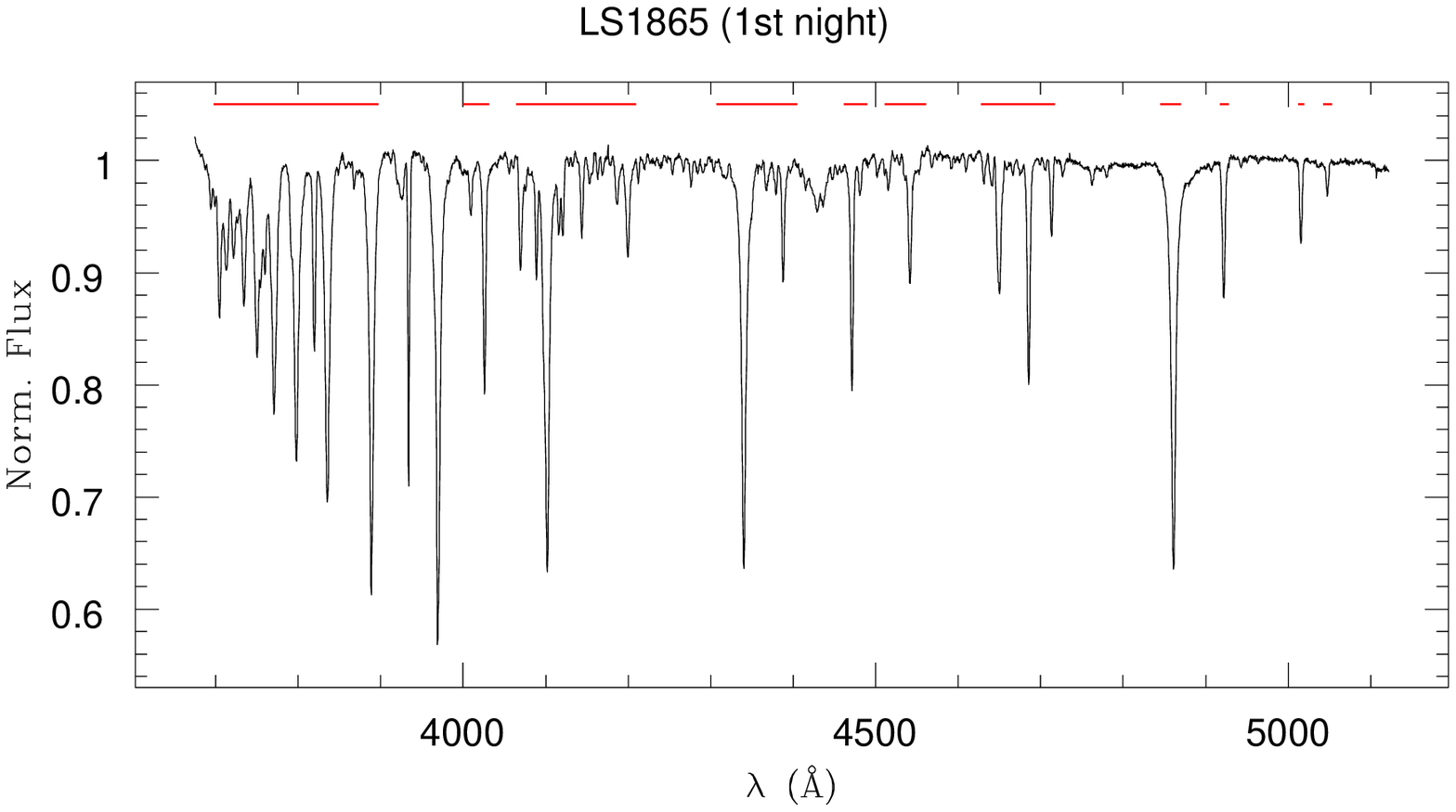}
\caption{(Roughly) normalized spectra of the targets, with spectral windows used for the magnetic field evaluation shown as thick red lines above the spectra.}
\label{spec}
\end{figure*}

\setcounter{figure}{0}
\begin{figure*}
\includegraphics[width=8cm, bb=60 430 565 720, clip]{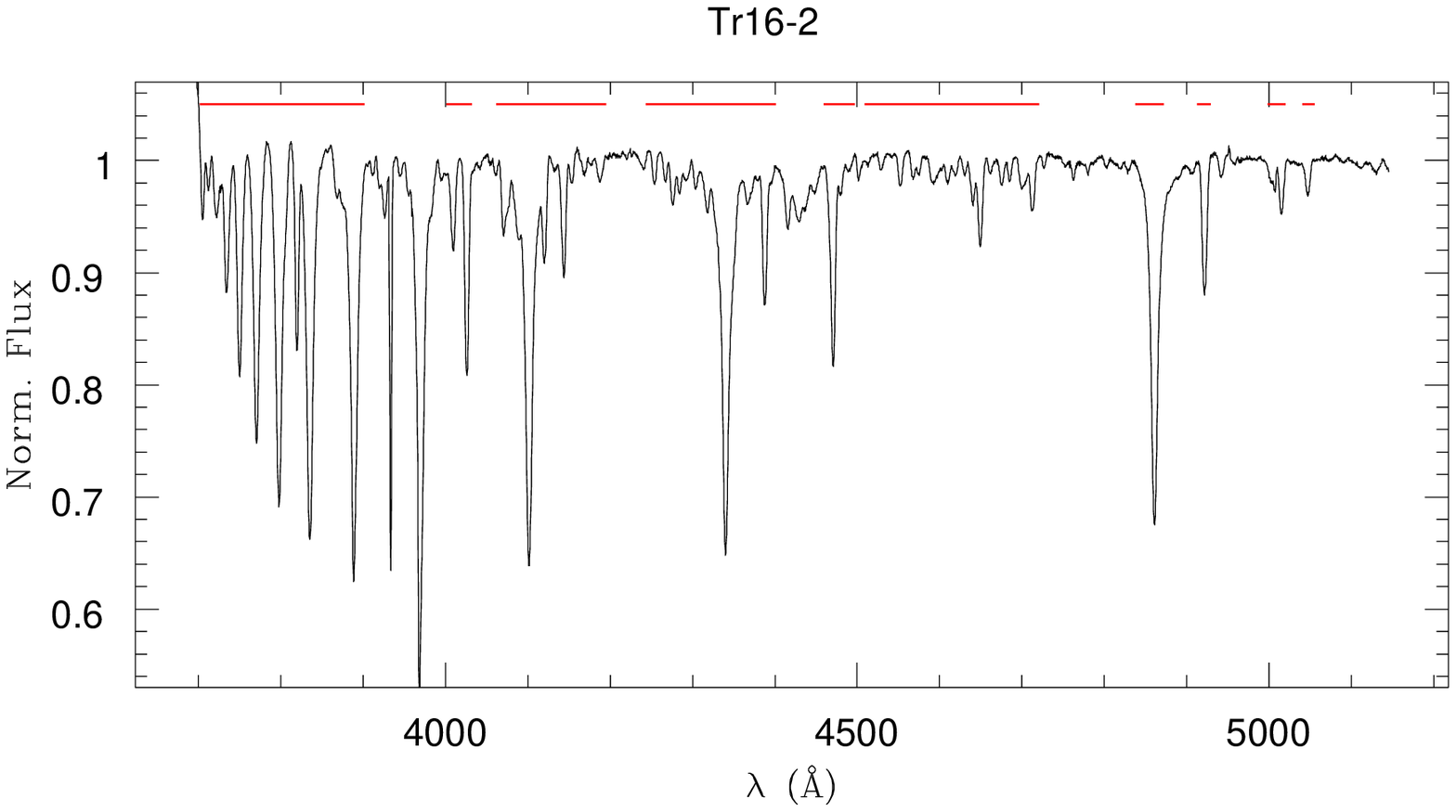}
\includegraphics[width=8cm, bb=60 430 565 720, clip]{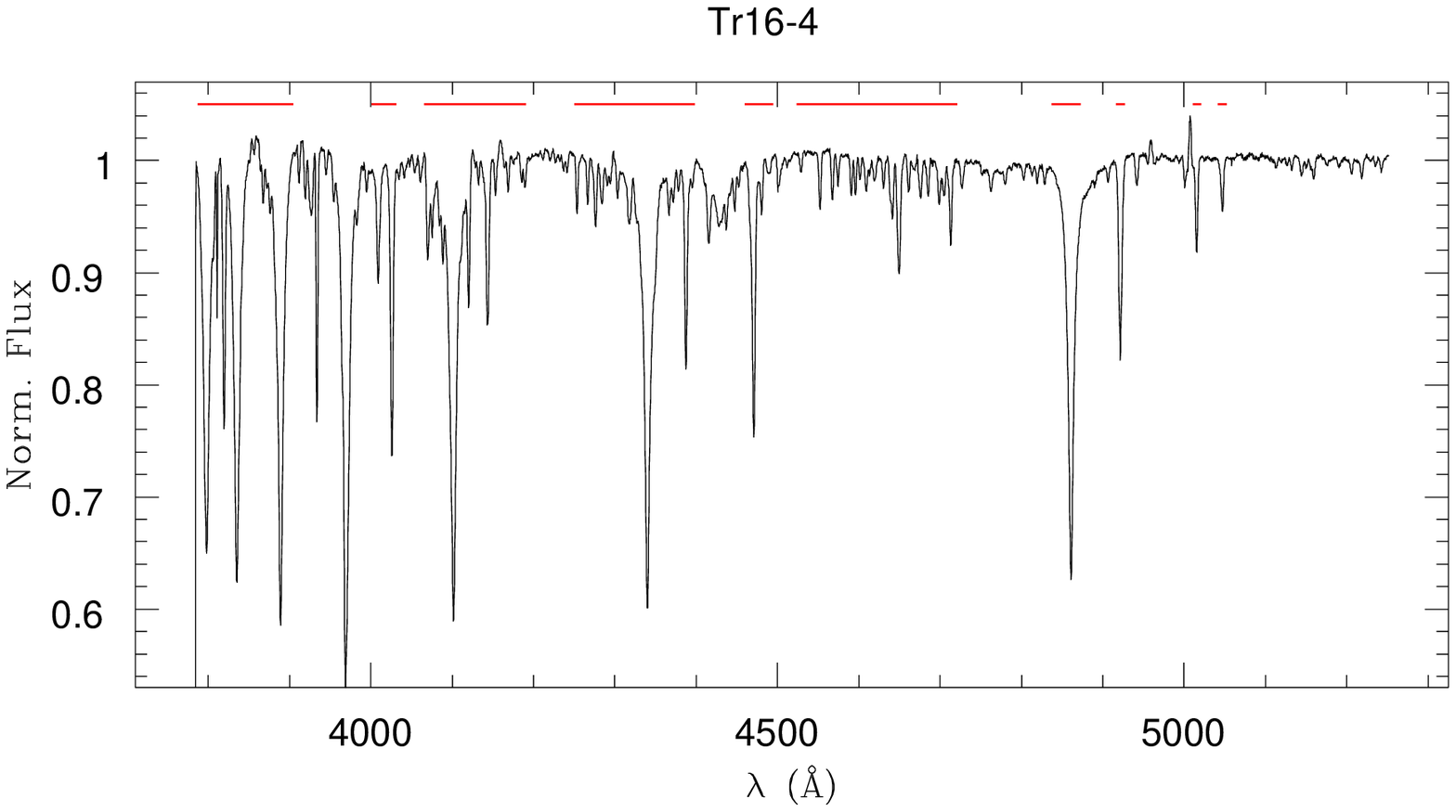}
\includegraphics[width=8cm, bb=60 430 565 720, clip]{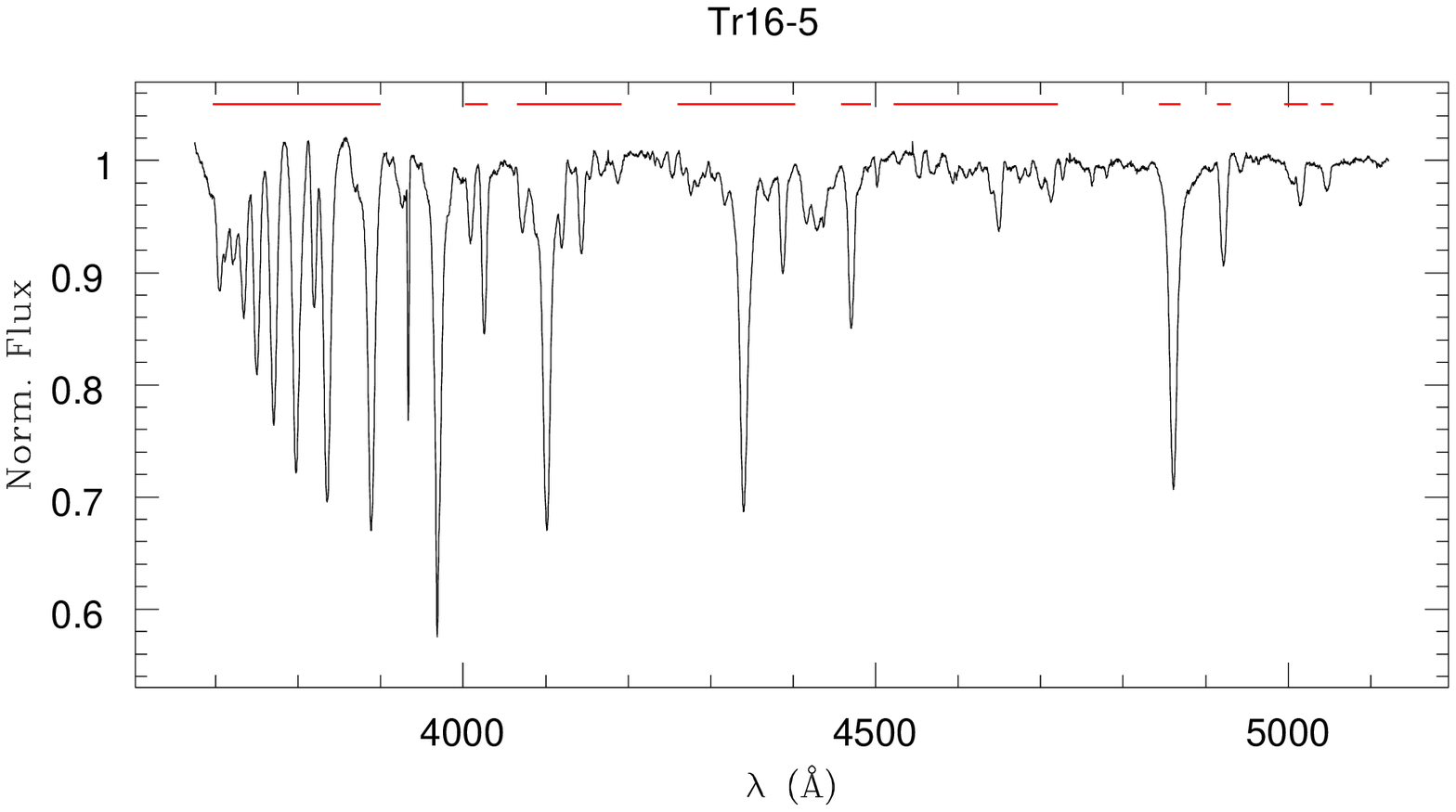}
\includegraphics[width=8cm, bb=60 430 565 720, clip]{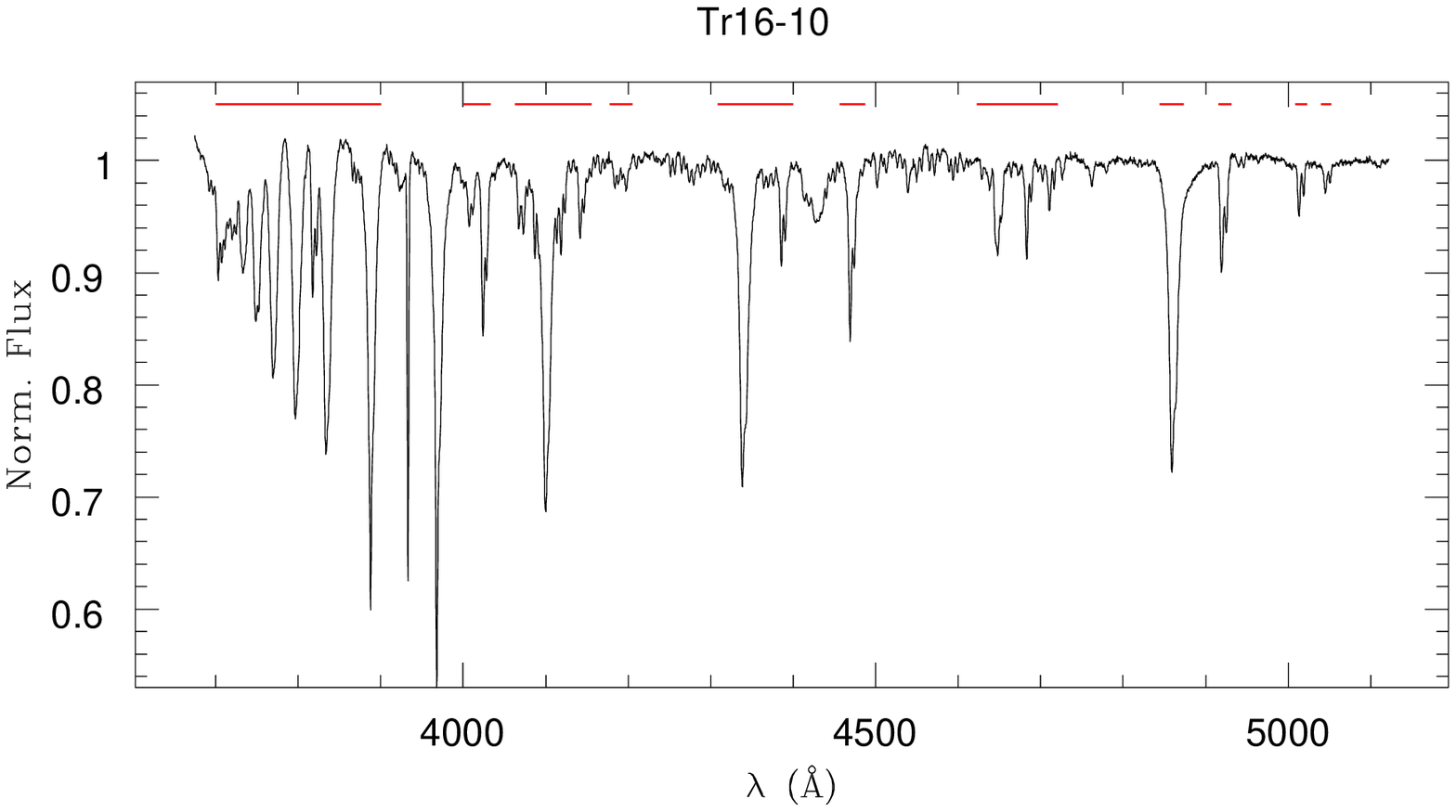}
\includegraphics[width=8cm, bb=60 430 565 720, clip]{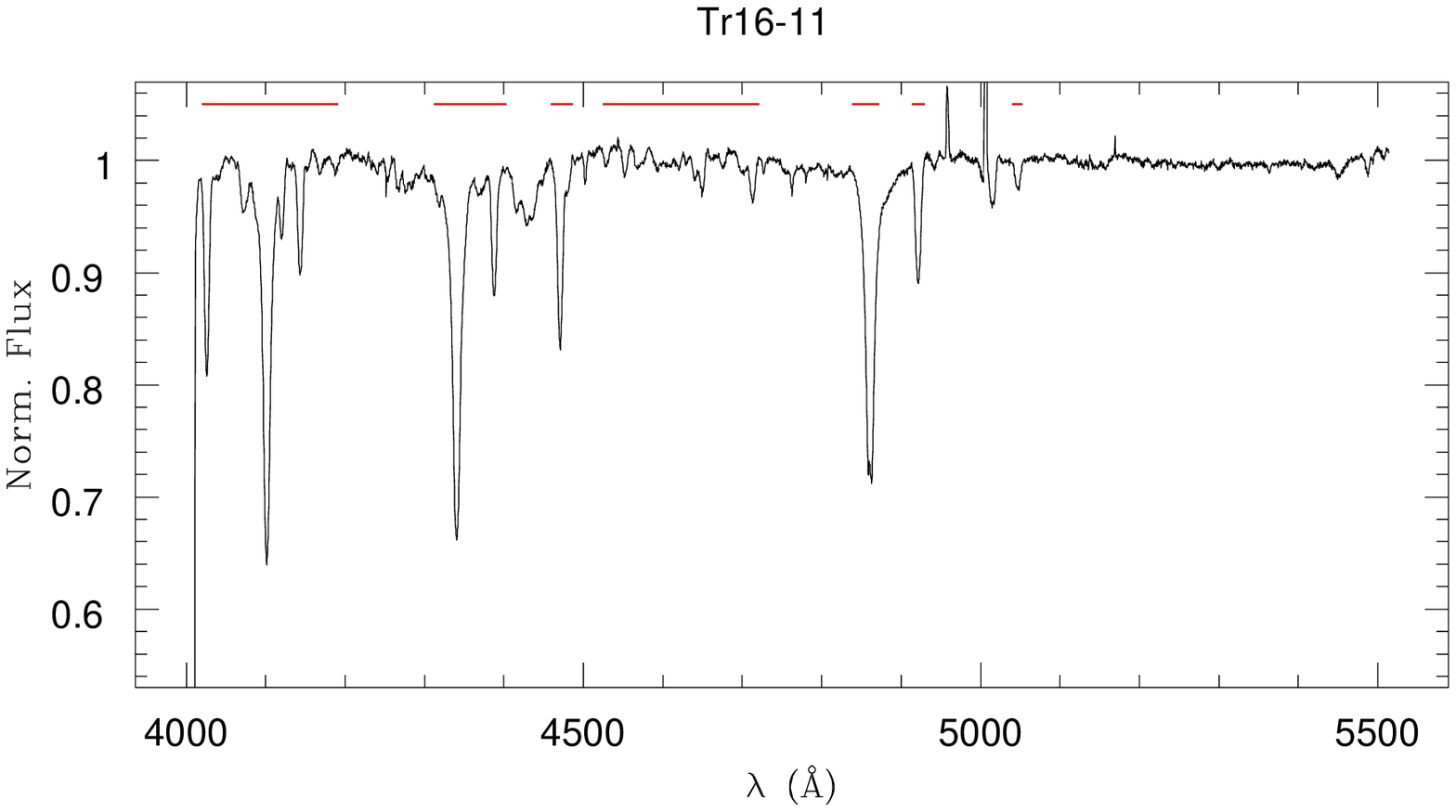}
\includegraphics[width=8cm, bb=60 430 565 720, clip]{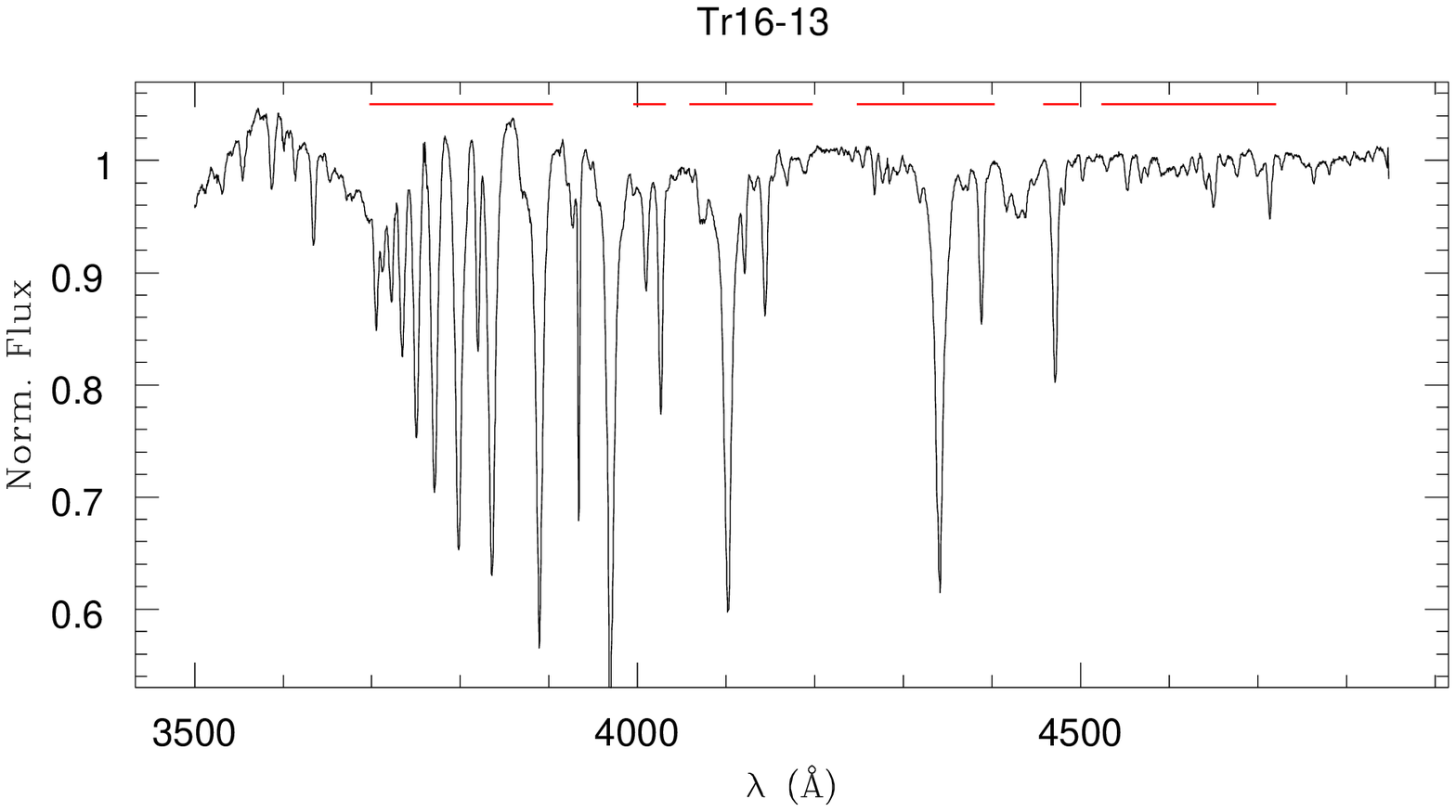}
\includegraphics[width=8cm, bb=60 430 565 720, clip]{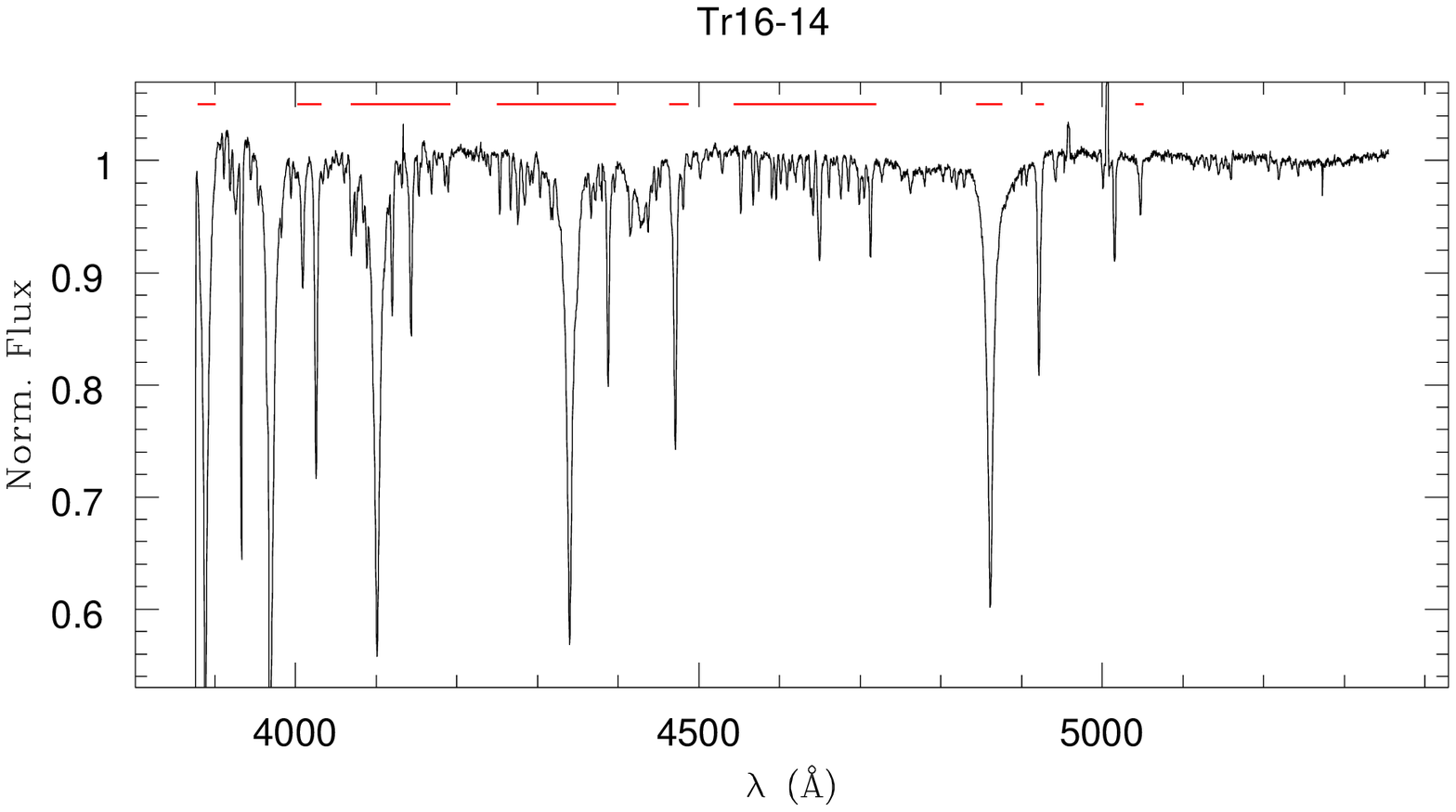}
\includegraphics[width=8cm, bb=60 430 565 720, clip]{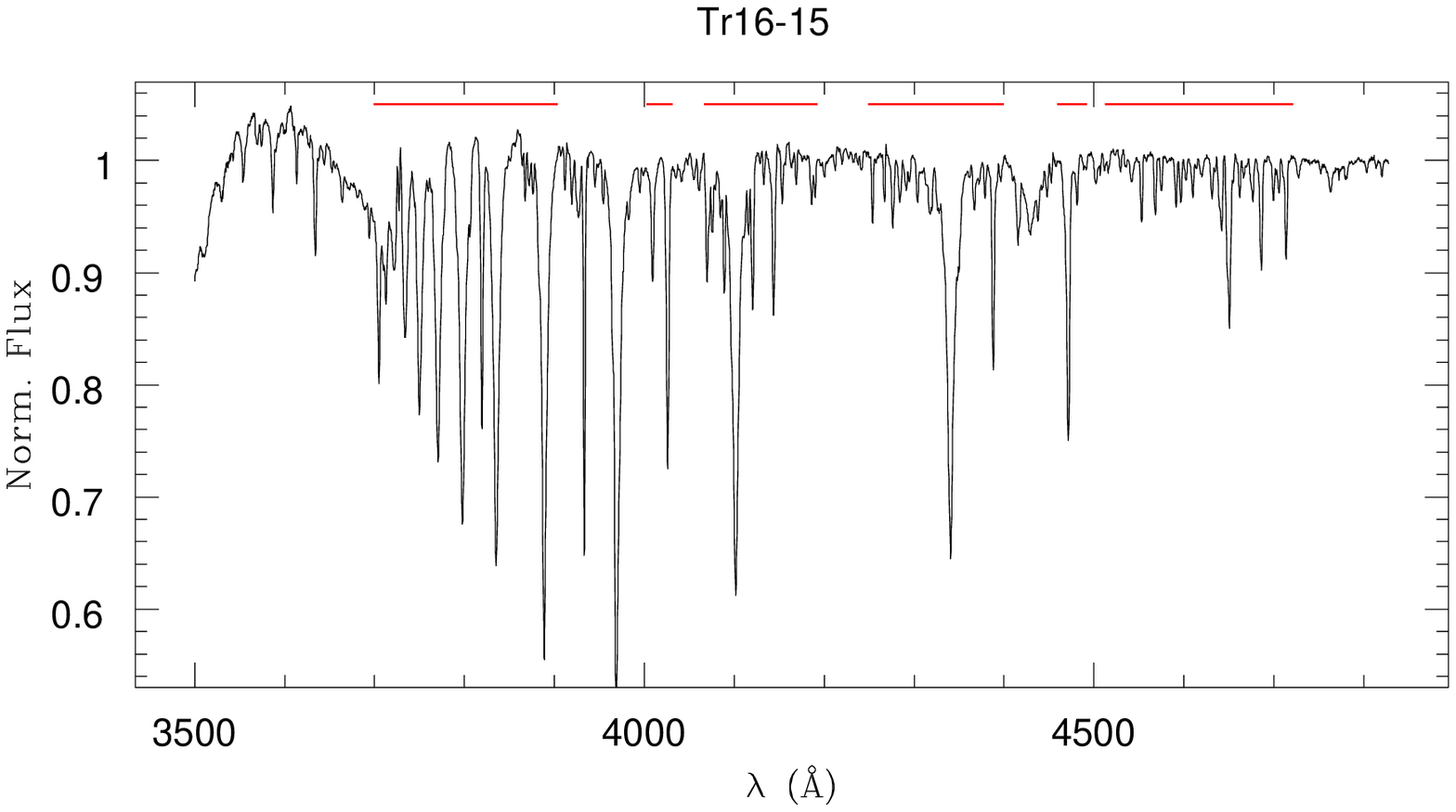}
\includegraphics[width=8cm, bb=60 430 565 720, clip]{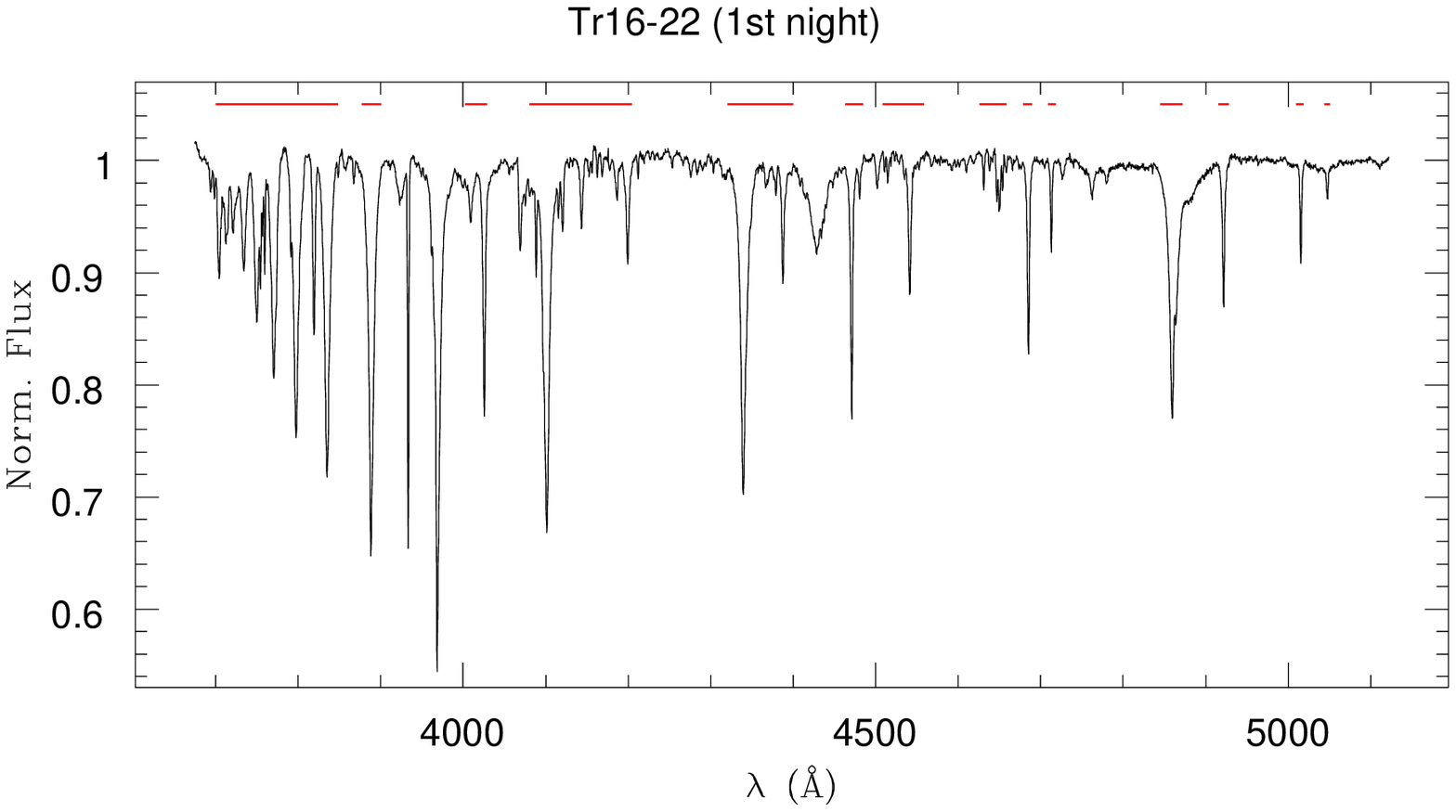}
\includegraphics[width=8cm, bb=60 430 565 720, clip]{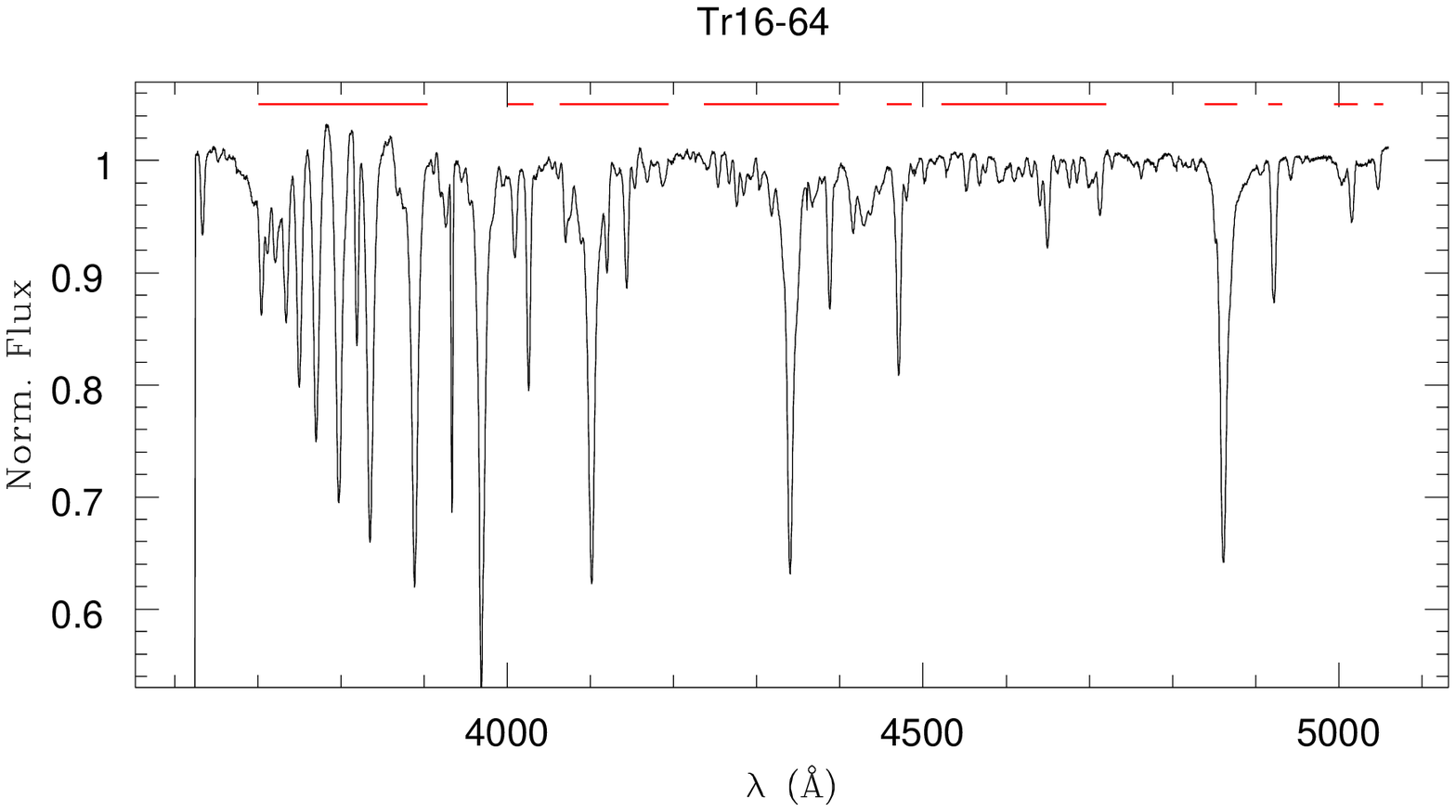}
\caption{ - Continued}
%\label{spec}
\end{figure*}

\setcounter{figure}{0}
\begin{figure*}
\includegraphics[width=8cm, bb=60 430 565 720, clip]{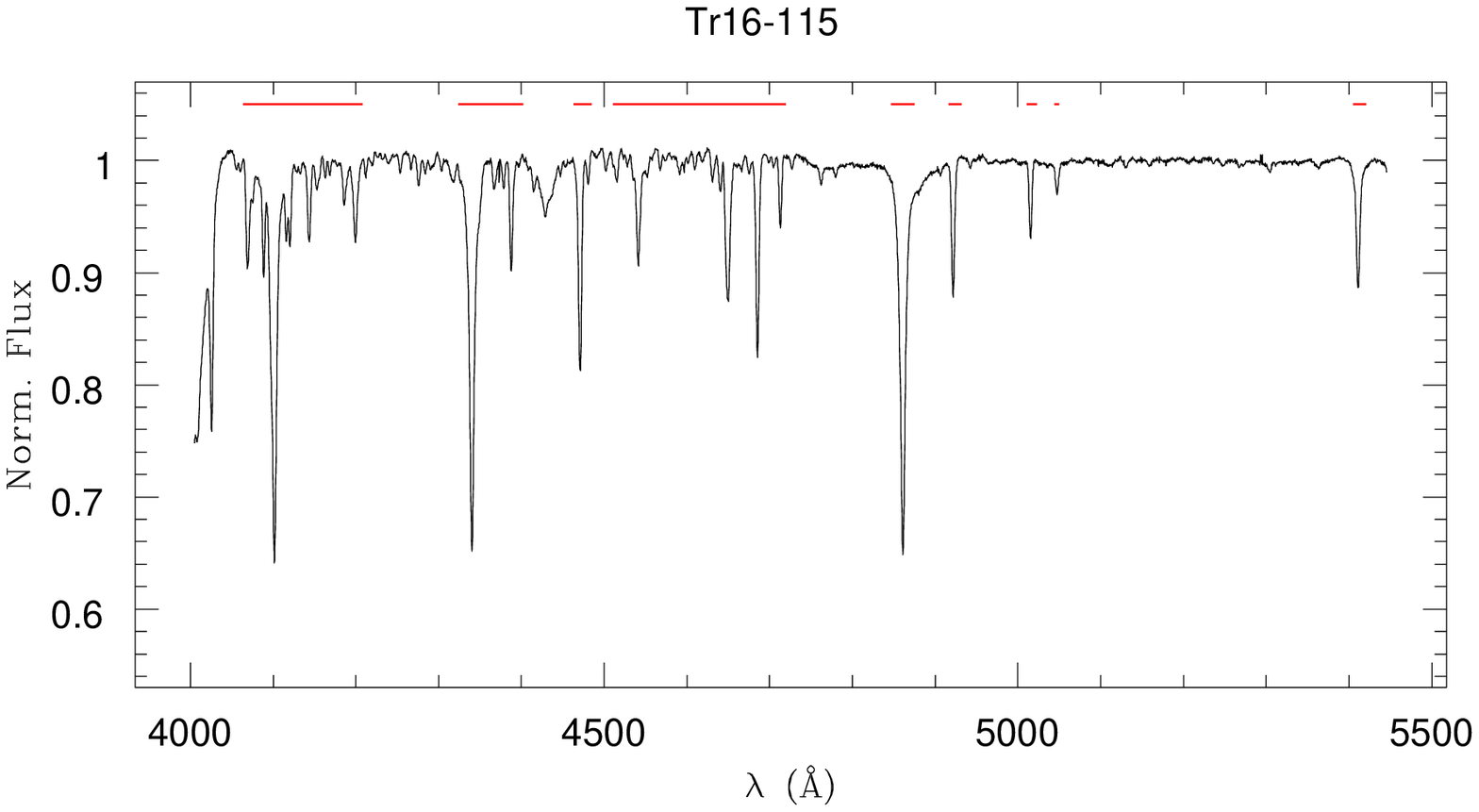}
\caption{ - Continued}
%\label{spec}
\end{figure*}

\label{lastpage}


\begin{thebibliography}{99}
\bibitem[\protect\citeauthoryear{Appenzeller et al.}{1998}]{app98} Appenzeller, I., Fricke, K., F{\"u}rtig, W., et al.\ 1998, The Messenger, 94, 1 

%\bibitem[\protect\citeauthoryear{Baade}{1991}]{baa91} Baade, D.\ 1991, European Southern Observatory Conference and Workshop Proceedings, 36, 21 

%\bibitem[\protect\citeauthoryear{Balona}{1992}]{bal92} Balona, L.~A.\ 1992, MNRAS, 254, 404 

\bibitem[\protect\citeauthoryear{Antokhin et al.}{2008}]{ant08} Antokhin, I.~I., Rauw, G., Vreux, J.-M., van der Hucht, K.~A., \& Brown, J.~C.\ 2008, A\&A, 477, 593 

\bibitem[\protect\citeauthoryear{Babel \& Montmerle}{1997a}]{bab97a} Babel, J., \& Montmerle, T.\ 1997, A\&A, 323, 121 

\bibitem[\protect\citeauthoryear{Babel \& Montmerle}{1997b}]{bab97b} Babel, J., \& Montmerle, T.\ 1997, ApJ, 485, L29 

\bibitem[\protect\citeauthoryear{Bagnulo et al.}{2012}]{bag12} Bagnulo, S., Landstreet, J.~D., Fossati, L., \& Kochukhov, O.\ 2012, A\&A, 538, A129 

\bibitem[\protect\citeauthoryear{Bagnulo et al.}{2009}]{bag09} Bagnulo, S., Landolfi, 
M., Landstreet, J.~D., et al.\ 2009, PASP, 121, 993 

\bibitem[\protect\citeauthoryear{Bagnulo et al.}{2002}]{bag02} Bagnulo, S., Szeifert, T., Wade, G.~A., Landstreet, J.~D., \& Mathys, G.\ 2002, A\&A, 389, 191 

\bibitem[\protect\citeauthoryear{Bell}{2004}]{bel04} Bell, A.~R.\ 2004, MNRAS, 353, 550 

\bibitem[\protect\citeauthoryear{Broos et al.}{2011}]{bro11} Broos, P.~S., Townsley, L.~K., Feigelson, E.~D., et al.\ 2011, ApJS, 194, 2 

\bibitem[\protect\citeauthoryear{Combi et al.}{2011}]{com11} Combi, J.~A., Albacete-Colombo, J.~F., Luque Escamilla, P.~L., et al.\ 2011, Bulletin de la Societe Royale des Sciences de Liege (BSRSL), 80, 644 

\bibitem[\protect\citeauthoryear{Donati et al.}{1997}]{don97} Donati, J.-F., Semel, M., Carter, B.~D., Rees, D.~E., \& Collier Cameron, A.\ 1997, MNRAS, 291, 658 

\bibitem[\protect\citeauthoryear{Evans et al.}{2004}]{eva04} Evans, N.~R., Schlegel, E.~M., Waldron, W.~L., et al.\ 2004, ApJ, 612, 1065 

\bibitem[\protect\citeauthoryear{Falceta-Gon{\c c}alves \& Abraham}{2012}]{gon12} Falceta-Gon{\c c}alves, D., \& Abraham, Z.\ 2012, MNRAS, in press (arXiv:1203.5093)


\bibitem[\protect\citeauthoryear{Gagn{\'e} et al.}{2005}]{gag05} Gagn{\'e}, M., Oksala, M.~E., Cohen, D.~H., et al.\ 2005, ApJ, 634, 712 (see also erratum in ApJ, 628, 986)


%\bibitem[\protect\citeauthoryear{Gagn{\'e} et al.}{2005}]{2005ApJ...628..986G} Gagn{\'e}, M., Oksala, M.~E., Cohen, D.~H., et al.\ 2005, ApJ, 628, 986 


\bibitem[\protect\citeauthoryear{Grunhut et al.}{2009}]{gru09} Grunhut, J.~H., Wade, G.~A., Marcolino, W.~L.~F., et al.\ 2009, MNRAS, 400, L94 


\bibitem[\protect\citeauthoryear{Grunhut et al.}{2012}]{gru12} Grunhut, J.~H., Rivinius, T., Wade, G.~A., et al.\ 2012, MNRAS, 419, 1610 


\bibitem[\protect\citeauthoryear{G{\"u}del \& Naz{\'e}}{2009}]{gud09} G{\"u}del, M., \& Naz{\'e}, Y.\ 2009, A\&AR, 17, 309 

%\bibitem[\protect\citeauthoryear{Hubrig et al.}{2008}]{hub08} Hubrig, S., Sch{\"o}ller, M., Schnerr, R.~S., et al.\ 2008, A\&A, 490, 793 


\bibitem[\protect\citeauthoryear{Hubrig et al.}{2011}]{hub11} Hubrig, S., Sch{\"o}ller, M., Kharchenko, N.~V., et al.\ 2011, A\&A, 528, A151 


\bibitem[\protect\citeauthoryear{Leitherer et al.}{1995}]{lei95} Leitherer, C., Chapman, J.~M., \& Koribalski, B.\ 1995, ApJ, 450, 289 

\bibitem[\protect\citeauthoryear{Maeder \& Meynet}{2003}]{mae03} Maeder, A., \& Meynet, G.\ 2003, A\&A, 411, 543 


\bibitem[\protect\citeauthoryear{Naz{\'e}}{2009}]{naz09} Naz{\'e}, Y.\ 2009, A\&A, 506, 1055 

\bibitem[\protect\citeauthoryear{Naz{\'e} et al.}{2010}]{naz10} Naz{\'e}, Y., Ud-Doula, A., Spano, M., et al.\ 2010, A\&A, 520, A59 

\bibitem[\protect\citeauthoryear{Naz{\'e} et al.}{2011}]{naz11} Naz{\'e}, Y., Broos, P.~S., Oskinova, L., et al.\ 2011, ApJS, 194, 7 

\bibitem[\protect\citeauthoryear{Naz{\'e} et al.}{2012}]{naz12} Naz{\'e}, Y., Zhekov, S.~A., \& Walborn, N.~R.\ 2012, ApJ, 746, 142 

\bibitem[\protect\citeauthoryear{Oskinova et al.}{2011}]{osk11} Oskinova, L.~M., Todt, H., Ignace, R., et al.\ 2011, MNRAS, 416, 1456 

\bibitem[\protect\citeauthoryear{Petit et al.}{2011}]{pet11} Petit, V., Massa, D.~L., Marcolino, W.~L.~F., et al.\ 2011, MNRAS, 412, L45 


\bibitem[\protect\citeauthoryear{Rauw et al.}{2009}]{rau09} Rauw, G., Naz{\'e}, Y., Fern{\'a}ndez Laj{\'u}s, E., Lanotte, A.~A., Solivella, G.~R., Sana, H., \& Gosset, E.\ 2009, MNRAS, 398, 1582 

\bibitem[\protect\citeauthoryear{Rivinius et al.}{2010}]{riv10} Rivinius, T., Szeifert, T., Barrera, L., et al.\ 2010, MNRAS, 405, L46 

\bibitem[\protect\citeauthoryear{Sana et al.}{2011}]{san11} Sana, H., Le Bouquin, J.-B., De Becker, M., et al.\ 2011, ApJ, 740, L43 

\bibitem[\protect\citeauthoryear{Stahl et al.}{1996}]{sta96} Stahl, O., Kaufer, A., Rivinius, T., et al.\ 1996, A\&A, 312, 539 

\bibitem[\protect\citeauthoryear{Townsend \& Owocki}{2005}]{tow05} Townsend, R.~H.~D., \& Owocki, S.~P.\ 2005, MNRAS, 357, 251 


\bibitem[\protect\citeauthoryear{Townsley et al.}{2011}]{tow11} Townsley, L.~K., Broos, P.~S., Corcoran, M.~F., et al.\ 2011, ApJS, 194, 1 

\bibitem[\protect\citeauthoryear{ud-Doula \& Owocki}{2002}]{udd02} ud-Doula, A., \& Owocki, S.~P.\ 2002, ApJ, 576, 413 

\bibitem[\protect\citeauthoryear{ud-Doula et al.}{2006}]{udd06} ud-Doula, A., Townsend, R.~H.~D., \& Owocki, S.~P.\ 2006, ApJ, 640, L191 

\bibitem[\protect\citeauthoryear{ud-Doula et al.}{2009}]{udd09} ud-Doula, A., Owocki, S.~P., \& Townsend, R.~H.~D.\ 2009, MNRAS, 392, 1022 

\bibitem[\protect\citeauthoryear{van Loo et al.}{2005}]{van} van Loo, S., Runacres, M.~C., \& Blomme, R.\ 2005, A\&A, 433, 313 
\bibitem[\protect\citeauthoryear{Williams et al.}{2011}]{wil11} Williams, S.~J., Gies, D.~R., Hillwig, T.~C., McSwain, M.~V., \& Huang, W.\ 2011, AJ, 142, 146 

\end{thebibliography}
\end{document}